\documentclass[12pt, a4paper]{article}
\usepackage{amsmath,amssymb}

\usepackage{graphicx}
\usepackage{multirow}
\usepackage{subfigure}
\usepackage{bm}
\usepackage{cite}
\usepackage{url}
\usepackage{array}
\usepackage{epstopdf}

\begin{document}

\title{Affective Music Information Retrieval}

\author{{Ju-Chiang Wang,$^*$ Yi-Hsuan Yang, and Hsin-Min Wang} \\ \\
        \normalsize{Institute of Information Science,}\\
        \normalsize{Academia Sinica, Taipei, Taiwan}\\
        \normalsize{$^*$ Corresponding e-mail: \texttt{asriver.wang@gmail.com}}}

\maketitle

\begin{abstract}
Much of the appeal of music lies in its power to convey emotions/moods and to evoke them in listeners. In consequence, the past decade witnessed a growing interest in modeling emotions from musical signals in the music information retrieval (MIR) community. In this article, we present a novel generative approach to music emotion modeling, with a specific focus on the valence-arousal (VA) dimension model of emotion. The presented generative model, called \emph{acoustic emotion Gaussians} (AEG), better accounts for the subjectivity of emotion perception by the use of probability distributions. Specifically, it learns from the emotion annotations of multiple subjects a Gaussian mixture model in the VA space with prior constraints on the corresponding acoustic features of the training music pieces. Such a computational framework is technically sound, capable of learning in an online fashion, and thus applicable to a variety of applications, including user-independent (general) and user-dependent (personalized) emotion recognition and emotion-based music retrieval. We report evaluations of the aforementioned applications of AEG on a larger-scale emotion-annotated corpora, AMG1608, to demonstrate the effectiveness of AEG and to showcase how evaluations are conducted for research on emotion-based MIR. Directions of future work are also discussed.
\end{abstract}

\section{Introduction}
\label{sec: INTRO}

Automatic music emotion recognition (MER) aims at modeling the association between music and emotion so as to facilitate emotion-based music organization, indexing, and retrieval. This technology has emerged in recent years as a promising solution to deal with the huge amount of music information available digitally \cite{kim10ismir,huq10jnmr,yang12tist,barthet12cmmr}.
It is generally believed that music cannot be composed, performed, or listened to without affection involvement \cite{emotion_book}.
The pursuit of emotional experience has also been identified as one of the primary motivations and benefits of music listening \cite{juslin04jnmr}.
In addition to music retrieval, music emotion also finds applications in context-aware music recommendation, playlist generation, music therapy, and automatic music accompaniment for other media content, including image, video, and text, amongst others \cite{lonsdale11bjp,wang12grand,yang13tmm,saari13ismir}. 

Despite of the significant progress that has been made in recent years, MER is still considered as a challenging problem because the perception of emotion in music is usually highly subjective.
A single, static ground-truth emotion label is not sufficient to describe the possible emotions different people perceive in the same piece of music \cite{gabrielsson02,huron06book}.
On the contrary, it may be more reasonable to learn a computational model from multiple responses of different listeners \cite{raykar10jmlr} and to present \emph{probabilistic} (soft) rather than \emph{deterministic} (hard) emotion assignments as the final result.
In addition, the subjective nature of emotion perception suggests the need of personalization in systems for emotion-based music recommendation or retrieval \cite{yang09sigir}.
Early work on MER often chose to sidestep this critical issue by either assuming that a common consensus can be achieved \cite{wang04icsp,huq10jnmr}, or by simply discarding music pieces for which a common consensus cannot be achieved \cite{lu06taslp}.

To help address this issue, we have proposed a novel generative model referred to as \emph{acoustic emotion Gaussians} (AEG) in our prior work \cite{wang12acmmm,wang12apsipa,wang12grand,wang12mirum,wang15tac}. The name of the AEG model comes from its use of multiple Gaussian distributions to model the affective content of music. The algorithmic part of AEG has been first introduced in \cite{wang12acmmm}, along with the preliminary evaluation of AEG for MER and emotion-based music retrieval. More details about the analysis part of the model learning of AEG can be found in a recent article \cite{wang15tac}. Due to the parametric nature of AEG, model adaptation techniques have also been proposed to personalize an AEG model in an online, incremental fashion, rather than learning from scratch \cite{wang12apsipa,chen14icassp}.
The goal of this article is to position the AEG model as a theoretical framework and to provide detailed information about the model itself and its application to personalized MER and emotion-based music retrieval.

We conceptualize emotion by the valence-arousal (VA) model \cite{russell80}, which has been used extensively by psychologists to study the relationship between music and emotion \cite{schubert04,eerola14aes}. 
These two dimensions are found to be the most fundamental through factor analysis of self-report of human's affective response to music stimulus. Despite differences in nomenclature, existing studies give similar interpretations of the resulting factors, most of which correspond to \emph{valence} (or pleasantness; positive/negative affective states) and \emph{arousal} (or activation; energy and stimulation level). For example, happiness is an emotion associated with a positive valence and a high arousal, while sadness is an emotion associated with a negative valence and a low arousal. We refer to the 2-D space spanned by valence and arousal as the \emph{VA space} hereafter.
Moreover, we are concerned with the emotion an individual \emph{perceives} as being expressed in a piece of music, rather than the emotion the individual actually \emph{feels} in response to the piece. This distinction is necessary \cite{gabrielsson02}, as we do not necessarily feel sorrow when listening to a sad tune, for example.

As the focus of this article is on \emph{dimensional} emotion values such as valence and arousal values, we refer interested readers to \cite{mirex07,schuller10jnmr,barthet12cmmr} for studies and surveys on \emph{categorical} MER research that views emotions as discrete labels such as mood tags. 
We also note that people have proposed approaches to model the relationship between discrete emotion labels and the dimensional VA values \cite{wang12mirum,saari14tkde}, which is also beyond the scope of this article.

The article is organized as follows. We first review related work in Section \ref{sec: RELATEDWORK}. Then, we present the mathematical derivation of AEG and the learning algorithm in Section \ref{sec: LEARING}, followed by the personalization algorithm in Section \ref{sec: PERSON}. Sections \ref{sec: MER} and \ref{sec: EMO_MR} present applications of AEG to MER and emotion-based music retrieval, respectively. Finally, we conclude in Section \ref{sec: CONCLUSION}.

\section{Related Work on Dimensional Music Emotion Recognition}
\label{sec: RELATEDWORK}

Early approaches to MER \cite{karl07jnmr,yang08taslp} assumed that the perceived emotion of a music piece can be represented as a
\emph{single point} in the VA space, in which the valence and arousal values are considered as independent numerical values. The ground-truth VA
values of a music piece is obtained by averaging the annotations of a number of human subjects, without considering the covariance
of the annotations. To predict the VA values of a music piece, a regression model can be applied. Given $N$ inputs $(\mathbf{x}_i, y_i)$, $i=1,\ldots,N$, where $\mathbf{x}_i$ is a $D$-dimensional feature vector of the $i$-th input segment, $D$ the number of feature descriptors,
and $y_i$ the valence or arousal value, a regression model is learned by algorithms such as support vector regression (SVR)
\cite{ML_nu_SVR} that minimize the mismatch (e.g. mean squared loss) between the predicted and the ground-truth VA values.

As emotion perception is rarely dependent on a single music factor but a combination of them \cite{hevner35,juslin00}, algorithms
used feature descriptors that characterize the loudness, timbre, pitch, rhythm, melody, harmony or lyrics of music \cite{hu10ismir,schuller10jnmr,panda13cmmr,schmidt13ismir}. In particular, while it is usually easier to predict arousal using,
for example, loudness and timbre features, the prediction of valence has been found more challenging \cite{schuller10jnmr,yang11taslpA,wang13aam}. Cross cultural aspects of emotion perception have also been studied \cite{hu14smc}. 
To exploit the temporal continuity of emotion variation within a piece of music, techniques such as system identification \cite{korhonen06tsmc}, conditional random fields \cite{schmidt11ismir,imbrasaite13aam}, hidden Markov models \cite{madsen14ismir}, deep recurrent neural networks \cite{weninger14icassp}, or dynamic probabilistic model \cite{wang15icassp} have also been proposed. Various approaches and features for MER have been evaluated and compared using benchmarking datasets comprising over 1,000 Creative Commons licensed music pieces from the Free Music Archive, in the 2013 and 2014 MediaEval `Emotion in Music' tasks  \cite{soleymani13crowdmm,soleymani14mm}.

\begin{figure}[!t]
\centering
\includegraphics[width=.9\columnwidth]{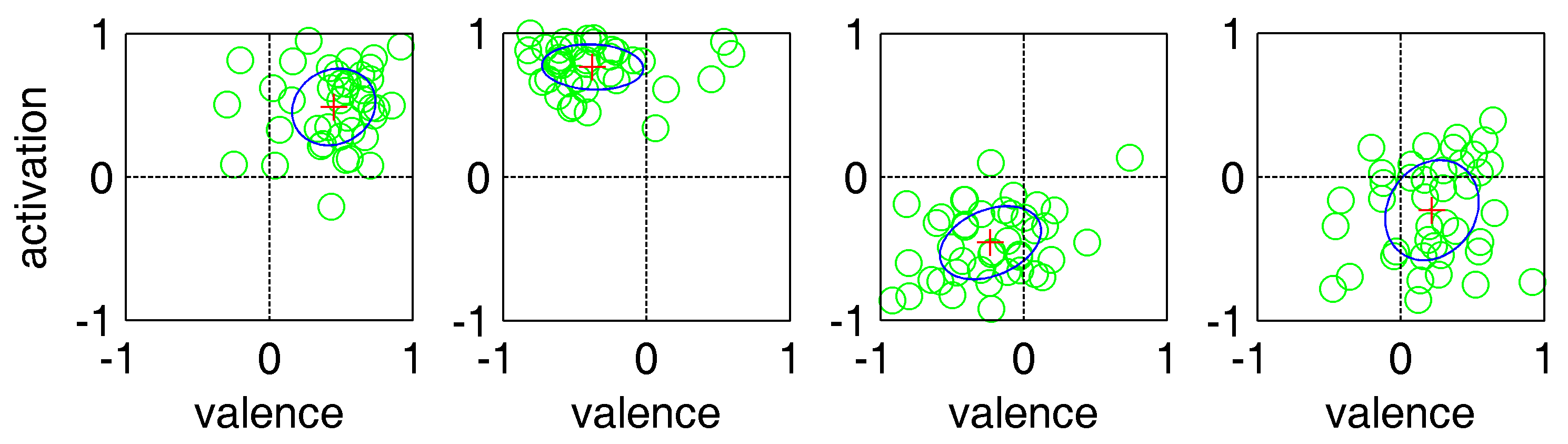}
\caption[]{Subjects' annotations of the perceived emotion of four 30-second clips, which from left to right are \emph{Dancing Queen} by ABBA, \emph{Civil War} by Guns N' Roses, \emph{Suzanne} by Leonard Cohen, and \emph{All I Have To Do Is Dream} by the Everly Brothers. Each circle here corresponds to a subject's annotation, and the overall emotion for a clip can be approximated by a 2-D Gaussian distribution (the red cross and blue ellipse). Note that throughout this article we use the contour of an ellipse to outline the standard deviation of the corresponding Gaussian distribution.}
\label{fig: EMO_DISTRI}
\end{figure}

Recent years have witnessed growing attempts to model the emotion of a music piece as a probability distribution in the VA space \cite{yang11taslpB,schmidt10ismir,wang12acmmm,chen14icassp} to better account for the subjective nature of emotion perception. For instance, Figure \ref{fig: EMO_DISTRI} shows the VA values applied by different annotators to four music pieces. To characterize the distribution of the emotion annotations for each clip, a typical way is to use a bivariate Gaussian distribution, where the mean vector presents the most possible VA values and the covariance matrix indicates its uncertainty. For a clip with highly subjective affective content, the determinant of the covariance matrix would be larger.

Existing approaches to predicting the emotion distribution of a music clip from acoustic features fall into two categories. The \emph{heatmap} approach \cite{yang11taslpB,schmidt11ismir} quantizes each emotion dimension by $W$ equally spaced cells, leading to a $W \times W$ grid representation of the VA space. The approach trains $W^2$ regression models for predicting the emotion \emph{intensity} of each cell. Higher intensity at a cell indicates that people are more likely to perceive the corresponding emotion from the clip. The emotion intensity over the VA space creates a heatmap-like representation of emotion distribution. However, heatmap is not a continuous representation of emotion and emotion intensity cannot be strictly considered as a probability estimate.

The \emph{Gaussian-parameter} approach \cite{yang11taslpB,schmidt10ismir}, on the other hand, models emotion distribution of a clip as a bivariate Gaussian and trains multiple regressors, each for a parameter of the mean vector and the covariance matrix. This makes it easy to apply lessons learned from modeling the mean VA values. In addition, performance analysis of this approach is easier; one can analyze the importance of different acoustic features to each Gaussian parameter individually. However, since the regression models are trained independently, the correlation between valence and arousal is not exploited. The parameter estimation of the mean and variance is disjoined as well.

A different methodology to address the subjectivity is to call for a user-dependent model trained on annotations of a specific user to personalize the emotion prediction \cite{yeh06pcm,yang07hcm,zhu08edutainment}. In \cite{yang07hcm}, two personalization methods are proposed; the first trains a \emph{personalized} MER system for each individual specifically, whereas the second groups users according to some personal factors (e.g. gender, music experience, and personality) and then trains \emph{group-wise} MER system for each user group. Another \emph{two-stage} personalization scheme has also been studied \cite{yang09sigir}: the first stage estimates the general perception of a music piece, whereas the second one predicts the difference between the general perception and the personal one of the target user.

We note that none of the aforementioned approaches renders a strict probabilistic interpretation \cite{wang15tac}. In addition, many existing work is developed on discriminative models such as multiple linear regression and SVR.
Few attempts are made to develop a principled probabilistic framework that is technically sound for modeling the music emotion and that permits  extending the user-independent model to a user-dependent one, preferably in an online fashion.

We also note that most existing work focuses on the \emph{annotation} aspect of music emotion research, namely MER. Little work has been made to the \emph{retrieval} aspect -- the development of emotion-based music retrieval systems \cite{yang12tist}. In what follows, we present the AEG model and its applications to the both of these two aspects.

\section{Acoustic emotion Gaussians: A Generative Approach for Music Emotion Modeling}
\label{sec: LEARING}

\begin{figure}[!t]
\centering
\includegraphics[width=\columnwidth]{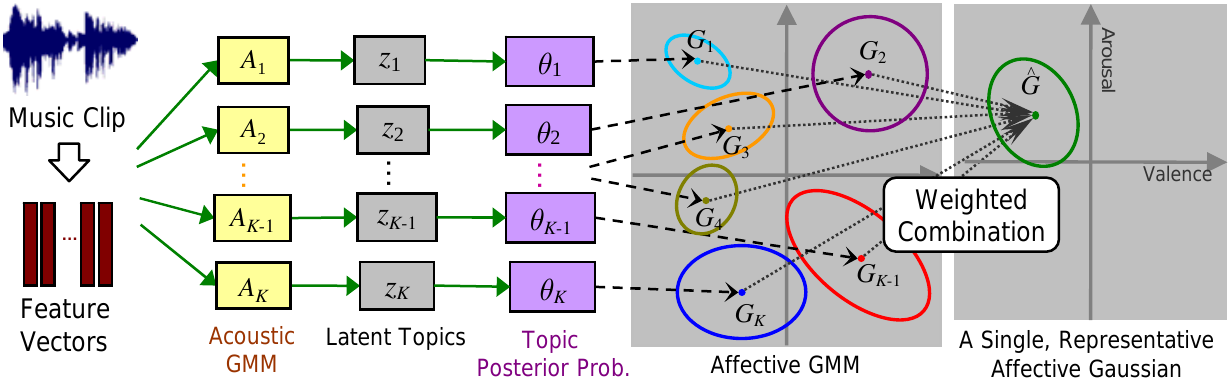}
\caption{Illustration of the generative process of the AEG model.}
\label{fig: EMO_SYS1}
\end{figure}

In \cite{wang12acmmm,wang12apsipa,wang12grand,wang12mirum,wang15tac}, we proposed AEG, which is fundamentally different from the existing regression or heatmap approaches. As Figure \ref{fig: EMO_SYS1} shows, AEG involves the generative process of VA emotion distributions from audio signals. While the relationship between audio and music emotion may sometimes be complicated and difficult to observe directly from an emotion-annotated corpus, AEG uses a set of clip-level \emph{latent topics} $\{z_k\}_{k=1}^K$ to resolve this issue.

We first define the terminology and explain the basic principle of AEG.
Suppose that there are $K$ \emph{audio descriptors} $\{A_k\}_{k=1}^K$, each is related to some acoustic feature vectors of music clips. Then, we map the associated feature vectors of $A_k$ to a clip-level topic $z_k$. To implement each $A_k$, we use a single Gaussian distribution in the acoustic feature space. The aggregated Gaussians of $\{A_k\}_{k=1}^K$ is called an \emph{acoustic GMM} (Gaussian mixture model). Subsequently, we map each $z_k$ to a specific area in the VA space, which is modeled by a bivariate Gaussian distribution $G_k$. We refer to the aggregated Gaussians of $\{G_k\}_{k=1}^K$ as an \emph{affective GMM}.
Given a clip, its feature vectors are first used to compute the posterior distribution over the topics, termed as a \emph{topic posterior representation} $\bm{\theta}$. In $\bm{\theta}$, the posterior probability of $z_k$ (denoted as $\theta_k$) is associated with $A_k$ and will then be used to show the clip's importance to $G_k$. Consequently, the posterior distribution $\bm{\theta}=\{\theta_k\}_{k=1}^K$ can be incorporated into learning the affective GMM as well as making emotion prediction for a clip.

AEG-based MER follows the flow depicted in Figure \ref{fig: EMO_SYS1}. Based on $\bm{\theta}$ of a test clip, we obtain the weighted affective GMM $\sum_k \theta_k G_k$, which is able to generate various emotion distribution. Following this sense, if a clip's acoustic features can be completely described by the $h$-th topic $z_h$, i.e. $\theta_h =1$, and $\theta_k=0$, $\forall k\neq h$, then its emotion distribution would exactly follow $G_h$. As will be described in Section \ref{sec: MER}, we can further approximate $\sum_k \theta_k G_k$ by a single, representative affective Gaussian $\hat G$ for simplicity. This is illustrated in the rightmost of Figure \ref{fig: EMO_SYS1}.

Beyond valence and arousal, adding more dimensions (e.g. \emph{potency}, or dominant--submissive) might help resolve the ambiguity between affective terms, such as anger and fear, which are close to one another in the second quadrant of the VA space \cite{bigand05,collier07}. Although AEG can be easily extended to describe emotion in higher dimensions, we stay with the 2-D emotion model here again for simplicity. 

\subsection{Topic Posterior Representation}
\label{sec: topic_post}

The topic posterior representation of a music clip is generated from its audio. We note that the temporal dynamics of audio signals is regarded as essential for human to perceive musical characteristics such as timbre, rhythm, and tonality. To capture more local temporal variation of the low-level features, we represent the acoustic features at a time instance in the segment-level, which corresponds to sufficiently long duration (e.g. 0.4 second). A segment-level feature vector $\mathbf{x}$ can be formed by, for example, concatenating the mean and standard deviation of the frame-level feature vectors within the segment. As a result, a clip is divided into multiple overlapped segments which are then represented by a sequence of vectors, $\{\mathbf{x}_1, \ldots, \mathbf{x}_{T}\}$, where $T$ is the length of the clip.

To start the generative process of AEG, we first learn an acoustic GMM as the bases to represent a clip. This acoustic GMM can be trained using the expectation-maximization (EM) algorithm on a large set of segment-level vectors $\mathcal{F}$ extracted from existing music clips. The learned acoustic GMM defines the set of audio descriptors $\{A_k\}_{k=1}^K$, and can be expressed as follows,
\begin{equation}
p(\mathbf{x}) = \sum_{k = 1}^K {\pi_k  A_k (\mathbf{x} \mid  {\mathbf{m}}_k ,{\mathbf{S}}_k )}\,,
\label{eq: p(x)}
\end{equation}
where $A_k(\cdot)$ is the $k$-th component Gaussian distribution, and $\pi_k$, $\mathbf{m}_k$, and $\mathbf{S}_k$ are its corresponding prior weight, mean vector, and covariance matrix, respectively.
Note that we substitute equal weight for the GMM (i.e. $\pi_k = \frac{1}{K}$, $\forall k$), because the original $\pi_k$ learned from $\mathcal{F}$ does not imply the prior distribution of the feature vectors in a clip. Such a heuristic usually results in better performance as pointed in \cite{wang11ismir}.

Suppose that we have an emotion annotated corpus $\mathcal{X}$ consisting of $N$ music clips $\{s_i\}_{i=1}^N$. Given a clip $s_i = \{\mathbf{x}_{i,t}\}_{t=1}^{T_i}$, we then compute the segment-level posterior probability for each feature vector in $s_i$ based on the acoustic GMM,
\begin{equation}
p(A_k \mid \mathbf{x}_{i,t} ) = \frac{ A_k (\mathbf{x}_{i,t}\mid  {\mathbf{m}}_k ,{\mathbf{S}}_k )} {\sum\nolimits_{h = 1}^K  A_h (\mathbf{x}_{i,t} \mid  {\mathbf{m}}_h ,{\mathbf{S}}_h ) }\,.
\label{eq: p(z|x)}
\end{equation}
Finally, the clip-level topic posterior probability $\theta_{i,k}$ of $s_i$ can be approximated by averaging the segment-level ones,
\begin{equation}
\theta_{i,k}  \leftarrow p(z_k \mid s_i ) \approx \frac{1}{{T_i }}\sum_{t = 1}^{T_i} p(A_k \mid \mathbf{x}_{i,t}) \,.
\label{eq: theta_ik}
\end{equation}
This approximation assumes that $\theta_{i,k}$ is equally contributed by each segment of $s_i$ and thereby capable of representing the clip's acoustic features. We use a vector $\bm{\theta}_i \in \mathbb{R}^K$, whose $k$-th component is $\theta_{i,k}$, as the topic posterior of $s_i$.

\subsection{Prior Model for Emotion Annotation}
\label{sec: PRI-ANO}

To consider the subjectivity of emotional responses of a music clip, we ask multiple subjects to annotate the clip. However, as some subjects' annotations may not be reliable, we introduce a \emph{user prior model} to quantify the contribution of each subject.

Let $\mathbf{e}_{i,j} \in \mathbb{R}^2$ (a vector including the valence and arousal values) denote one of the annotations of $s_i$ given by the $j$-th subject, and let $U_i$ denote the number of subjects who have annotated $s_i$. Note that $\mathbf{e}_{q,j}$ and $\mathbf{e}_{r,j}$, where $q\neq r$, may not correspond to the same subject. Then, we build the user prior model $\gamma$ to describe the confidence of $\mathbf{e}_{i,j}$ in $s_i$ using a single Gaussian distribution,
\begin{equation}
\gamma(\mathbf{e}_{i,j} \mid s_i ) \equiv G({\mathbf{e}}_{i,j} \mid  \mathbf{a}_i ,\mathbf{B}_i ),
\label{eq: gamma}
\end{equation}
where $\mathbf{a}_i=\frac{1}{U_i}\sum_{j=i}^{U_i} \mathbf{e}_{i,j}$, $\mathbf{B}_i = \frac{1}{U_i} \sum_{j=1}^{U_i} (\mathbf{e}_{i,j}-\mathbf{a}_i)(\mathbf{e}_{i,j}-\mathbf{a}_i)^T$, and $G(\mathbf{e} \mid \mathbf{a}_i, \mathbf{B}_i)$ is called the \emph{annotation Gaussian} of $s_i$. One can observe what $\mathbf{a}_i$ and $\mathbf{B}_i$ look like from the four example clips in Figure \ref{fig: EMO_DISTRI}. Empirical results show that a single Gaussian performs better than a GMM for setting up $\gamma(\cdot)$ \cite{wang12acmmm}.

The confidence of $\mathbf{e}_{i,j}$ can be estimated based on the likelihood calculated by Eq. \ref{eq: gamma}. If an annotation is far away from the mean, it gives small likelihood accordingly. In addition to Gaussian distributions, any criterion that is able to reflect the importance of a user's annotation of a clip can be applied to $\gamma$.

The probability of $\mathbf{e}_{i,j}$, referred to as the \emph{clip-level annotation prior}, can be calculated by normalizing the likelihood of $\mathbf{e}_{i,j}$ over the cumulative likelihood of all other annotations in $s_i$,
\begin{equation}
p(\mathbf{e}_{i,j} \mid s_i ) \equiv \frac{ \gamma ( \mathbf{e}_{i,j} \mid  s_i )} {\sum\nolimits_{r=1}^{U_i} \gamma (\mathbf{e}_{i,r} \mid  s_i ) }\,.
\label{eq: p(u|s)}
\end{equation}
Based on the clip-level annotation prior, we further define the \emph{corpus-level clip prior} to describe the importance of each clip,
\begin{equation}
p(s_i \mid \mathcal{X}) \equiv \frac{\sum\nolimits_{j=1}^{U_i} \gamma(\mathbf{e}_{i,j} \mid s_i)}
{ \sum\nolimits_{q=1}^N \sum\nolimits_{r=1}^{U_q} \gamma(\mathbf{e}_{q,r} \mid s_q) }\,.
\label{eq: p(s|C)}
\end{equation}
From Eqs. \ref{eq: p(u|s)} and \ref{eq: p(s|C)} we can make two observations. First, if a clip's annotations are consistent (i.e. $\mathbf{B}_i$ is small), it is considered less subjective. Second, if a clip is annotated by more subjects, the corresponding $\gamma$ model should be more reliable.
As a result, we can define the \emph{corpus-level annotation prior} $\gamma_{i,j}$ for each $\mathbf{e}_{i,j}$ in the corpus $\mathcal{X}$ by multiplying Eqs. \ref{eq: p(u|s)} and \ref{eq: p(s|C)}:
\begin{equation}
\gamma_{i,j} \leftarrow p(\mathbf{e}_{i,j} \mid \mathcal{X}) \equiv \frac{\gamma (\mathbf{e}_{i,j} \mid s_i)}
{\sum\nolimits_{q=1}^N \sum\nolimits_{r=1}^{U_q} \gamma(\mathbf{e}_{q,r} \mid s_i)}\,,
\label{eq: gamma2}
\end{equation}
which is computed beforehand and fixed in learning the affective GMM.


\subsection{Learning the Affective GMM}
\label{sec: AEG-LEARN}

Given a training music clip $s_i$ in the corpus $\mathcal{X}$, we assume the emotional responses can be generated from an affective GMM weighted by its topic posterior $\bm{\theta}_i$,
\begin{equation}
p(\mathbf{e}_{i,j} \mid \bm{\theta}_i) = \sum_{k = 1}^K {\theta_{i,k} G_k(\mathbf{e}_{i,j} \mid  \bm{\mu}_k ,\bm{\Sigma}_k )} \,,
\label{eq: p(e|s,theta)}
\end{equation}
where $G_k(\cdot)$ is the $k$-th affective Gaussian with mean $\bm{\mu}_k$ and covariance $\bm{\Sigma}_k$ to be learned. Here $\theta_{i,k}$ stands for the fixed weight associated with $A_k$ to carry the audio characteristics of $s_i$. We therefore call $\bm{\theta}_i$ an \emph{acoustic prior}.
Then, the objective function is in the form of the marginal likelihood function of the annotations:
\begin{equation}
\begin{split}
p(\mathbf{E} \mid \mathcal{X}, \bm{\Lambda}) &
= \sum_{i=1}^N p(s_i \mid \mathcal{X}) \sum_{j=1}^{U_i} p(\mathbf{e}_{i,j} \mid s_i) p(\mathbf{e}_{i,j} \mid  \bm{\theta}_i, \bm{\Lambda})  \\
& = \sum_{i=1}^N \sum_{j=1}^{U_i} p(s_i \mid \mathcal{X})p(\mathbf{e}_{i,j} \mid s_i) p({\mathbf{e}}_{i,j} \mid  \bm{\theta}_i, \bm{\Lambda})  \\
& = \sum_{i=1}^N \sum_{j=1}^{U_i} p(\mathbf{e}_{i,j} \mid \mathcal{X}) \sum_{k=1}^K \theta_{i,k} G_k( \mathbf{e}_{i,j} \mid \bm\mu_k, \bm\Sigma_k ) \,,\\
\end{split}
\label{eq: p(E|C)}
\end{equation}
where $\mathbf{E} = \{\mathbf{e}_{i,j}\}_{i=1,j=1}^{N,U_i}$, $\mathcal{X} = \{s_i,\bm{\theta}_i\}_{i=1}^N$, and $\bm{\Lambda}=\{\bm{\mu}_k,\bm{\Sigma}_k\}_{k=1}^K$ is the parameter set of the affective GMM.
Taking the logarithm of Eq. \ref{eq: p(E|C)} and replacing $p(\mathbf{e}_{i,j} \mid \mathcal{X})$ by $\gamma _{i,j}$ leads to
\begin{equation}
L = \log \sum_i \sum_j \gamma_{i,j} \sum_k \theta _{i,k} G_k(\mathbf{e}_{i,j} \mid \bm{\mu}_k , \bm{\Sigma}_k) \,,
\label{eq: L}
\end{equation}
where ${\sum_i}{\sum_j}{\gamma_{i,j}}=1$. To learn the affective GMM, we can maximize the log-likelihood in Eq. \ref{eq: L} with respect to the Gaussian parameters.
We first derive a lower bound of $L$ according to Jensen's inequality,
\begin{equation}
L \ge L_\text{bound}=\sum_i \sum_j \gamma _{i,j} \log \sum_k \theta _{i,k} G_k(\mathbf{e}_{i,j} \mid \bm{\mu}_k, \bm{\Sigma}_k ) \,.
\label{eq: Lbound}
\end{equation}
Then, we treat $L_\text{bound}$ as a surrogate of $L$ and use the EM algorithm \cite{bishop2006} to estimate the parameters of the affective GMM.
In the E-step, we derive the expectation over the posterior distribution of $z_k$ for all the training annotations,
\begin{equation}
Q = \sum_i \sum_j  \gamma_{i,j} \sum_k  p(z_k \mid \mathbf{e}_{i,j}) \Big( \log \theta_{i,k} + \log G_k(\mathbf{e}_{i,j} \mid \bm{\mu}_k, \bm{\Sigma}_k ) \Big) \,,
\label{eq: Q}
\end{equation}
where
\begin{equation}
p(z_k \mid \mathbf{e}_{i,j}) = \frac{\theta _{i,k} G_k(\mathbf{e}_{i,j} \mid \bm{\mu}_k, \bm{\Sigma}_k )}
{\sum\nolimits_{h = 1}^K \theta _{i,h} G_k(\mathbf{e}_{i,j} \mid \bm{\mu}_h, \bm{\Sigma}_h)} \,.
\label{eq: p(z|e,theta)}
\end{equation}
In the M-step, we first set the derivative of Eq. \ref{eq: Q} with respect to $\bm{\mu}_k$ to zero and obtain the updating form for the mean vector,
\begin{equation}
\bm{\mu}'_k  \leftarrow \frac{\sum_i \sum_j \gamma _{i,j} p(z_k \mid \mathbf{e}_{i,j}) \mathbf{e}_{i,j}} {\sum_i \sum_j \gamma_{i,j} p(z_k \mid \mathbf{e}_{i,j})}\,.
\label{eq: mu_update}
\end{equation}
Following a similar line of reasoning, we obtain the update rule for $\bm{\Sigma}_k$:
\begin{equation}
\bm{\Sigma}'_k  \leftarrow
\frac{\sum_i\sum_j\gamma_{i,j}p(z_k\mid\mathbf{e}_{i,j})(\mathbf{e}_{i,j}-\bm{\mu}'_k)(\mathbf{e}_{i,j}-\bm{\mu}'_k )^T }
{\sum_i \sum_j \gamma_{i,j} p(z_k \mid \mathbf{e}_{i,j})} \,.
\label{eq: sigma_update}
\end{equation}

Theoretically, the EM algorithm iteratively maximizes the $L_{\text{bound}}$ value in Eq. \ref{eq: Lbound} until convergence. One can fix the number of maximal iterations or set a stopping criterion for the increasing ratio of $L_{\text{bound}}$.

Note that we can ignore the annotation prior by setting a uniform distribution, i.e., $\forall i, j$, $\gamma_{i,j} = 1$. This case is called ``AEG Uniform'' in the experiment. In contrast, the case with non-uniform annotation prior is called ``AEG AnnoPrior.''



\subsection{Discussion}

As Eqs. \ref{eq: mu_update} and \ref{eq: sigma_update} show, the re-estimated parameters $\bm{\mu}'_k$ and $\bm{\Sigma}'_k$ are collectively contributed by $\mathbf{e}_{i,j}, \forall~i,~j$, with the weights governed by the product of $\gamma_{i,j}$ and $p(z_k \mid  \mathbf{e}_{i,j})$. Consequently, the learning process seamlessly takes the annotation prior, acoustic prior, and annotation clusters over the current affective GMM into consideration. In such a way, the annotations of different clips can be shared with one another according to their corresponding prior probabilities. This can be a key factor that enables AEG to generalize the audio-to-emotion mapping.

As the affective GMM is getting fitted to the data, a small number of affective Gaussian components might overly fit to some emotion annotations, rendering the so-called \emph{singularity} problem \cite{bishop2006}. When this occurs, the corresponding covariance matrices would become non-positive definite (non-PD). Imagining that when a component affective Gaussian is contributed by only one or two annotations, the corresponding covariance shape will become a point or a straight line in the VA space. To tackle this issue, we can remove the component Gaussian when it happens to produce a non-PD covariance matrix during the EM iterations \cite{wang15tac}.

We note that ``early stop'' is a very important heuristic while learning the affective GMM. We find that setting a small number for the maximal iteration (e.g. 7 -- 11) or a larger stopping threshold for the increasing ratio of $L_{\text{bound}}$ (e.g. 0.01) empirically leads to better generalizability. It can not only prevent the aforementioned singularity problem but also avoid overly fitting to the training data. Empirical results show that the accuracy of MER improves as the iteration evolves and then degrades when the optimal iteration number has reached \cite{wang15tac}. Moreover, AEG AnnoPrior empirically converges faster and learns smaller covariances than AEG Uniform does.

\section{Personalization with AEG}
\label{sec: PERSON}

The capability for personalization is a very important characteristic that completes the AEG framework, making it more applicable to real-world applications. As AEG is a probabilistic, parametric model, it can incorporate personal information of a particular user via model adaptation techniques to make custom predictions.
While such personal information may include personal emotion annotation, user profile, transaction records, listening history, and relevance feedback, we focus on the use of personal emotion annotations in this article.

Because of the cognitive load for annotating music emotion, it is usually not easy to collect a sufficient amount of personal annotations at once to make the system reach an acceptable performance level. On the contrary, a user may provide annotations sporadically in different listening sessions. To this end, an online learning strategy \cite{bottou-98x} is desirable. When the annotations of a target user are scarce, a good online learning method needs to prevent over-fitting to the personal data in order to keep certain model generalizability. In other words, we cannot totally ignore the contributions of emotion perceptions from other users. Motivated by the Gaussian Mixture Model-Universal Background Model (GMM-UBM) speaker verification system \cite{reynolds2000dsp}, we first treat the affective GMM learned from broad subjects (called \emph{background users}) as a \emph{background (general) model}, and then employ a \emph{maximum a posteriori} (MAP)-based method \cite{gauvain1994tasp,reynolds2000dsp} to update the parameters of the background model using the personal annotations in an online manner. Theoretically, the resulting \emph{personalized model} will appropriately find a good trade-off between the target user's annotations and the background model.

\subsection{Model Adaptation}

In what follows, the acoustic GMM will stay fixed throughout the personalization process, since it is used as a reference model to represent the music audio. In contrast, the affective GMM is assumed to be learned on plenty of emotion annotations from quite a few subjects, so it possesses a sufficient representation (well-trained parameters) for user-independent (i.e. general) emotion perceptions. Our goal is to learn the personal perception with respect to the affective GMM $\bm{\Lambda}$ accordingly.

Suppose that we have a target user $u_\star$ annotating $M$ number of music clips denoted as $\mathcal{X_\star} = \{\mathbf{e}_i,\bm{\theta}_i\}_{i=1}^M$, where $\mathbf{e}_i$ and $\bm{\theta}_i$ are the emotion annotation and the topic posterior of a clip, respectively. We first compute each posterior probability over the latent topics based on the background affective GMM,
\begin{equation}
p(z_k \mid \mathbf{e}_i, \bm{\theta}_i ) = \frac{\theta_{i,k} G_k( \mathbf{e}_i \mid \bm{\mu}_k, \bm{\Sigma}_k )} {\sum_{h = 1}^K  \theta_{i,h} G_k( \mathbf{e}_i \mid \bm{\mu}_h, \bm{\Sigma}_h ) }.
\label{eq: ad_p(z|e,theta)}
\end{equation}
Then, we derive the expected sufficient statistics on $\mathcal{X_\star}$ over the posterior distribution of $p(z_k \mid \mathbf{e}_i, \bm{\theta}_i)$ for the mixture weight, mean, and covariance parameters:
\begin{align}
\Gamma_k = & \sum_{i=1}^M p( z_k \mid \mathbf{e}_i , \bm{\theta}_i )\,,
\label{eq: Ew} \\
\mathbb{E} (\bm{\mu}_k ) = & \frac{1}{\Gamma_k}\sum_{i=1}^M p( z_k \mid \mathbf{e}_i, \bm{\theta}_i) \mathbf{e}_i  \,,
\label{eq: Emuk} \\
\mathbb{E} (\bm{\Sigma}_k ) = & \frac{1}{\Gamma_k}\sum_{i=1}^M p( z_k \mid \mathbf{e}_i, \bm{\theta}_i )
\big(\mathbf{e}_i - \mathbb{E}(\bm{\mu}_k) \big) \big(\mathbf{e}_i -\mathbb{E} (\bm{\mu}_k) \big)^T \,.
\label{eq: ESigmak}
\end{align}
Finally, the new parameters of the personalized affective GMM can be obtained according to the MAP criterion \cite{gauvain1994tasp}. The resulting update rules are the forms of interpolations between the expected sufficient statistics (i.e. $E(\bm{\mu}_k)$ and $E(\bm{\Sigma}_k)$) and the parameters of the background model (i.e. $\bm{\mu}_k$ and $\bm{\Sigma}_k$) as follows:
\begin{equation}
\bm{\mu}'_k \leftarrow \alpha_k^\text{m} \mathbb{E}(\bm{\mu}_k) + \left( 1 - \alpha_k^\text{m} \right) {\bm{\mu}_k} \, ,
\label{eq: Amuk}
\end{equation}
\begin{equation}
\bm{\Sigma}'_k \leftarrow \alpha_k^\text{v} \mathbb{E}(\bm{\Sigma}_k) + \left( 1- \alpha_k^\text{v} \right) \left(\bm{\Sigma}_k + \bm{\mu}_k \bm{\mu}_k^T \right) - \bm{\mu}'_k ({\bm{\mu}'_k})^T\,.
\label{eq: ASigmak}
\end{equation}
The coefficients $\alpha_k^\text{m}$ and $\alpha_k^\text{v}$ are data-dependent and are defined as
\begin{equation}
\alpha_k^\text{m} = \frac{\Gamma_k} {\Gamma_k + \beta^\text{m}} \,,~~~ \alpha_k^\text{v} = \frac{\Gamma_k} {\Gamma_k + \beta^\text{v}} \,,
\label{eq: inter_alpha}
\end{equation}
where $\beta^\text{m}$ and $\beta^\text{v}$ are related to the hyper parameters \cite{gauvain1994tasp} and thus should be empirically defined by users. Note that there is no need to update the mixture weights, as they are already occupied by the fixed topic posterior weights.

\subsection{Discussion}

The MAP-based method is preferable in that we can determine the interpolation factor that balances the contribution between the personal annotations and the background model without loss of model generalizability, as demonstrated by its superior effectiveness and efficiency in speaker adaptation tasks \cite{reynolds2000dsp}. If a personal annotation $\{\mathbf{e}_m,\bm{\theta}_m\}$ is highly correlated to a latent topic $z_k$ (i.e. $p(z_k | \mathbf{e}_m, \bm{\theta}_m)$ is large), the annotation will contribute more to the update of $\{\bm{\mu}'_k, \bm{\Sigma}'_k\}$. In contrast, if the user's annotations have nothing to do with $z_h$ (i.e. the cumulative posterior probability $\Gamma_h=0$), the parameters of $\{\bm{\mu}'_h, \bm{\Sigma}'_h\}$ would remain the same as those of the background model, as shown by the fact that $\alpha_k$ would be 0.

Another advantage of the MAP-based method is that users are free to provide personal annotations for whatever songs they like, such as the songs they are more familiar with. This can help reduce the cognitive load of the personalization process. As the AEG framework is audio-based, the annotated clips can be arbitrary and does not have to be those included in the corpus for training the background model.

Finally, we note that the model adaptation procedure only needs to be performed once, so the algorithm is fairly efficient. It only requires $K$ times of computing the expected sufficient statistics and updating the parameters. In consequence, we can keep refining the background model whenever a small number of personal annotations are available, and readily use the updated model for personalized MER or music retrieval. The model adaptation method for GMM is not limited to the MAP method. We refer interested readers to \cite{chen14icassp, chen15taslp} for more advanced methods.

\section{AEG-based Music Emotion Recognition}
\label{sec: MER}

\subsection{Algorithm}
\label{sec: Emo_Predict}

As described in Section \ref{sec: LEARING}, we predict the emotion distribution of an unseen clip by weighting the affective GMM using the clip's topic posterior $\bm{\hat\theta} = \{\hat \theta_k\}_{k=1}^K$ as
\begin{equation}
p(\mathbf{e} \mid \bm{\hat\theta}) = \sum_{k = 1}^K \hat\theta_{k} G_k( \bm{\mu}_k ,\bm{\Sigma}_k ) \,.
\label{eq: predict_form}
\end{equation}
In addition, we can also use a single, representative affective Gaussian $G(\hat{\bm{\mu}}, \hat{\bm{\Sigma}})$ to summarize the weighted affective GMM. This can be done by solving the following optimization problem:
\begin{equation}
\underset{\bm{\hat\mu}, \bm{\hat\Sigma}}{\min} ~~ \sum_{k=1}^K \hat\theta_k D_\text{KL} \big( G_k( \bm\mu_k, \bm\Sigma_k ) ~\big|\hspace{-.45 mm}\big|~ G( \bm{\hat\mu}, \bm{\hat\Sigma} )  \big) \,,
\label{eq: gauss_merge}
\end{equation}
where
\begin{equation}
D_\text{KL} ( G_A \parallel  G_B) =
\frac{1}{2} \Big( \text{tr}(\bm{\Sigma}_A \bm{\Sigma}_B^{-1}) - \log \mid \bm{\Sigma}_A \bm{\Sigma}_B^{-1}\mid  + (\bm{\mu}_A-\bm{\mu}_B)^T {\bm{\Sigma }}_B^{-1} (\bm{\mu}_A-\bm{\mu}_B) - 2 \Big)
\label{eq: KL}
\end{equation}
denotes the one-way (asymmetric) Kullback--Leibler (KL) divergence (a.k.a. relative entropy) \cite{Kul51} from $G_A(\bm{\mu}_A, \bm{\Sigma}_A)$ to $G_B(\bm{\mu}_B, \bm{\Sigma}_B)$.
This optimization problem is strictly convex in $\bm{\hat\mu}$ and $\bm{\hat\Sigma}$, which means that there is a unique minimizer for the two variables, respectively \cite{davis06nips}. Let the partial derivative with respect to $\bm{\hat\mu}$ be 0, we have
\begin{equation}
\sum\nolimits_k \hat\theta_k ( 2 \bm{\hat\mu} - 2 \bm{\mu}_k) =0 \, .
\end{equation}
Given the fact that $\sum_k \hat \theta_k = 1$, we derive
\begin{equation}
\bm{\hat\mu} = \sum_{k=1}^K {\hat \theta_k} \bm{\mu}_k \,.
\label{eq: mu_opt}
\end{equation}
Setting the partial derivative with respect to $\bm{\Sigma}_k^{-1}$ to 0,
\begin{equation}
\sum\nolimits_k \hat\theta_k \left( {\bm{\Sigma}}_k - \bm{\hat\Sigma} + \left( {\bm{\mu}}_k - \bm{\hat\mu} \right) {\left( {\bm{\mu}}_k - \bm{\hat\mu} \right)}^T \right) = 0\,,
\end{equation}
we obtain the optimal covariance matrix by,
\begin{equation}
\bm{\hat \Sigma} = \sum_{k=1}^K {\hat \theta _k} \left( {\bm{\Sigma }_k  + \left({\bm{\mu }}_k  - \bm{\hat\mu} \right) \left({\bm{\mu }}_k  - \bm{\hat\mu} \right)^T } \right) \,.
\label{eq: sigma_opt}
\end{equation}

\subsection{Discussion}

Representing the predicted result as a single Gaussian is functionally necessary, because it is easier and more straightforward to interpret or visualize the emotion prediction to the users with only a single mean (center) and covariance (uncertainty).
However, this may run counter to the theoretical arguments given in favor of a GMM that permits emotion modeling in finer granularity. For instance, it is inadequate for the excerpts whose emotional responses are by nature bi-modal. We note that in applications such as emotion-based music retrieval (cf. Section \ref{sec: EMO_MR}) and music video generation \cite{wang12grand}, one can directly use the raw weighted GMM (i.e. Eq. \ref{eq: predict_form}) as the emotion index of a song in response to queries given in the VA space. We will detail this aspect later in Section \ref{sec: EMO_MR}.

The computation of Eqs. \ref{eq: mu_opt} and \ref{eq: sigma_opt} is quite efficient. The complexity depends mainly on $K$ and the number of frames $T$ of a clip: computing $\theta_k$ requires $KT$ operations (cf. Eq. \ref{eq: p(z|x)}), whereas computing $\bm{\hat\mu}$ and $\bm{\hat\Sigma}$ requires $K$ vector multiplications and $K$ matrix operations, respectively.
This efficiency is important for dealing with a large-scale music database and for application such as real-time music emotion tracking on a mobile device \cite{schmidt11ismir,imbrasaite13aam,wang12icassp,wang14icme,wang15icassp}.

\subsection{Evaluation on General MER}
\label{sec: EXP-MER}

\subsubsection{Dataset}

We use the AMG1608 dataset \cite{chen15icassp} for evaluating both general and personalized MER. The dataset contains 1,608 30-second music clips annotated by 665 subjects (345 are male; average age is 32.0$\pm$11.4) recruited mostly from the crowdsourcing platform Mechanical Turk \cite{Pao10}.
The subjects were asked to rate the VA values that best describe their general (instead of moment-to-moment) emotion perception of each clip via the internet. The VA values, which are real values ranging in between [--1,~1], are entered by clicking on the emotion space on a square interface panel. The subjects were instructed to rate the perceived rather than felt emotion. Each music clip was annotated by 15--32 subjects. Each subject annotated 12--924 clips, and 46 out of the 665 subjects annotated more than 150 music clips, making the dataset a useful corpus for research on MER personalization. The average Krippendorff's $\alpha$ across the music clips is 0.31 for valence and 0.46 for arousal, which are both in the range of fair agreement. Please refer to \cite{chen15icassp} for more details about this dataset.

\subsubsection{Acoustic Features}

As different emotion perceptions are usually associated with different patterns of features \cite{emotion_book2},
we use two toolboxes, MIRtoolbox \cite{lartillot07dafx} and YAAFE \cite{Mat10}, to extract four sets of frame-based features from audio signals, including MFCC-related features, tonal features, spectral features, and temporal features, as listed in Table \ref{tab:features}. We down-sample all the audio clips in AMG1608 at 22,050 Hz and normalize them to the same volume level. All the frame-based features are extracted with the same frame size of 50ms and 50\% hop size. Each dimension in the frame-based feature vectors is normalized to zero mean and unit standard deviation. We concatenate all the four sets of features for each frame, as this leads to better performance in acoustic modeling in our pilot study \cite{chen15taslp}. As a result, a frame-level feature vector contains 72 dimensions of features.

\begin{table}[!t]
\small
\centering
\caption{Frame-based acoustic features used in the evaluation.}
\label{tab:features}
\begin{tabular}{|l|c|p{9.8cm}|}
\hline
Feature  & Dim. & Description                                                                                                                                                                                              \\ \hline
\multirow{2}{*}{MFCCs}     & \multirow{2}{*}{40}  & 20 Mel-frequency cepstral coefficients and the first-order time differences \cite{Dav80}.                                                                                                                  \\ \hline
\multirow{2}{*}{Tonal}    & \multirow{2}{*}{17}  & Octave band signal intensity using a triangular octave filter bank and the ratio of these intensity values \cite{Mat10}.\\
\hline
\multirow{3}{*}{Spectral} & \multirow{3}{*}{11}  & Linear predictor coefficients that capture the spectral envelope of the audio signal \cite{Mak75}, spectral flux,\cite{Mat10} and spectral shape descriptors \cite{Pee04}.\\
\hline
\multirow{2}{*}{Temporal} & \multirow{2}{*}{4}   & Shape and statistics (centroid, spread, skewness, and kurtosis) \cite{Gil04}. \\
\hline\hline
\multirow{2}{*}{All}      & \multirow{2}{*}{72}  & Concatenation of all the four types of features mentioned above.\\
\hline
\end{tabular}
\end{table}

However, it does not make sense to analyze and predict the music emotion on a specific frame.
Instead of bag-of-frames approach \cite{wang11ismir,Li2014TMM}, we adopt the bag-of-segments approach for the topic posterior representation, because a segment is able to capture more local temporal variation of the low-level features. Our preliminary result has also confirmed this hypothesis.
To generate a segment-level feature vector representing a basic term in the bag-of-segments approach, we concatenate the mean and standard deviation of 16 consecutive frame-level feature vectors, leading to a 144-dimensional vector for a segment. The hop size for a segment is 4 frames. Given the acoustic GMM (cf. Eq. \ref{eq: p(x)}), we then follow Eqs. \ref{eq: p(z|x)} and \ref{eq: theta_ik} addressed in Section \ref{sec: topic_post} to compute the topic posterior vector of a music clip.


\subsubsection{Evaluation Metrics}

The accuracy of general MER is evaluated using 3 performance metrics: two-way KL divergence (KL2) \cite{Kul51}, Euclidean distance, and $R^2$ (also known as the coefficient of determination) \cite{Sen90}. The first two measure the distance between the prediction and the ground truth. The lower the value is, the better the performance.
KL2 considers the performance with respect to the bivariate Gaussian distribution of a chip, while the Euclidean distance is concerned with the VA mean only.
$R^2$ is also concerned with the VA mean only. In contrast to the distance measure, a high $R^2$ value is preferred.  Moreover, $R^2$ is computed separately for valence and arousal.

Specifically, we are given the distribution of the ground truth annotations $\mathcal{N}_i= G( \mathbf{a}_i,\mathbf{B}_i)$ (cf. Section \ref{sec: PRI-ANO}) and the predicted distribution of each test clip $\hat{\mathcal{N}_i}=G( \hat{\bm{\mu}}_i, \hat{\bm{\Sigma}}_i)$, both of which are modeled as a bivariate Gaussian distribution, where $i \in \{1,\ldots,N\}$ denotes the index of a clip in the test set.
Instead of one-way KL divergence (cf. Eq. \ref{eq: KL}) for determining the representative Gaussian, we evaluate the performance of emotion distribution prediction based on the KL2 divergence defined by
\begin{equation}
D_\text{KL2} (G_A, G_B) \equiv \frac{1}{2} \Big( D_\text{KL}(G_A \parallel G_B) + D_\text{KL}( G_B \parallel G_A) \Big) \,.
\label{eq: kl_two}
\end{equation}
The average KL2 divergence (AKL), which measures the symmetric distance between the predicted emotion distribution and the ground truth one, is computed by $\tfrac{1}{N}\sum\nolimits_{i=1}^{N} D_\text{KL2}(\mathcal{N}_i, \hat{\mathcal{N}_i})$.
Using the $l_2$ norm, we can compute the average Euclidean distance (AED) between the mean vectors of two Gaussian distributions by $\frac{1}{N} \sum\nolimits_{i=1}^{N}{\| \mathbf{a}_i - \bm{\hat\mu} \|_2 }$.
The $R^2$ statistics is a standard way to measure the fitness of regression models \cite{Sen90}.
It is used to evaluate the prediction accuracy as follows:
\begin{equation} \label{eq: r2}
R^2=
1-\frac{\sum_{i=1}^N (\hat{e_i} - e_i )^2 }{\sum_{i=1}^{N}{ (e_i - \bar{e} )^2 }
}\,,
\end{equation}
where $\hat{e_i}$ and $e_i$ denote the predicted (either valence or arousal) value and the ground truth one of a clip, respectively, and $\bar{e}$ is the average ground truth value over the test set. When the predictive model perfectly fits the ground truth values, $ R^2 $ is equal to $ 1 $. If the predictive model does not fit the ground truth well, $ R^2 $ may become negative.

We perform three-fold cross-validation to evaluate the performance of general MER. Specifically, the AMG1608 dataset is randomly partitioned into three folds, and an MER model is trained on two of them and tested on the other one.
Each round of validation generates the predicted result of one-third of the complete dataset. After three rounds, we will have the predicted result of each clip in the complete dataset. Then, AKL, AED, and the $R^2$ for valence and arousal are computed over the complete dataset, instead of computing the performance over each one-third of the dataset. This strategy gives an unbiased estimate for $R^2$.

\subsubsection{Result}
\label{sec: GMER_result}

We compare the performance of AEG with two baseline methods. The first one, referred to as the \emph{base-rate} method, uses a reference affective Gaussian whose mean and covariance are set using the global mean and covariance of the training annotations without taking into account the acoustic features. In other words, the prediction for every test clip would be the same for the base-rate method. The performance of this base-rate method can be considered as a lower bound in this task accordingly.
Moreover, we compare the performance of AEG with SVR \cite{ML_nu_SVR}, a representative regression-based approach for predicting emotion values or distributions, using the same type of acoustic features. Specifically, the feature vector of a clip is formed by concatenating the mean and standard deviation of all the frame-level feature vectors within a clip, yielding a 144-dimensional vector. We use the radial basis function (RBF) kernel SVR implemented by the libSVM library\cite{Cha11}, with parameters optimized by grid search with three-fold cross-validation on the training set. We further use a heuristic favorable for SVR to regularize every invalid predicted covariance parameter \cite{wang15tac}. This heuristic significantly improves the AKL performance of SVR.

Our pilot study empirically shows that AEG Uniform gives better emotion prediction in AED, compared to AEG AnnoPrior, possibly because the introduction of the annotation prior (cf. Eq. \ref{eq: gamma2}) may bias the estimation of the mean parameters in the EM learning. In contrast, AEG AnnoPrior leads to better result in AKL, indicating its capability of estimating a more proper covariance for a learned affective GMM. In light of this, we use a following \emph{hybrid} method to take advantage of both AEG AnnoPrior and AEG Uniform in optimizing the affective GMM. Suppose that we have learned two affective GMMs, one for AEG AnnoPrior and the other for AEG Uniform. To generate a combined affective GMM, for its $k$-th component Gaussian, we take the mean from the $k$-th Gaussian of AEG Uniform and the covariance from the $k$-th Gaussian of AEG AnnoPrior. This combined affective GMM is eventually used to predict the emotion for a test clip with Eqs. \ref{eq: mu_opt} and \ref{eq: sigma_opt} in this evaluation.


Table \ref{tab: mer_summary} compares the performance of AEG with the two baseline methods. It can be seen that both SVR and AEG outperform the base-rate method by a great margin, and that AEG can outperform SVR. For AEG, we can obtain better AKL and better $R^2$ for valence when $K=128$, but better AED and better $R^2$ for arousal when $K=256$. The best $R^2$ achieved for valence and arousal are 0.1601 and 0.6686. In particular, the superior performance of AEG in $R^2$ for valence is remarkable. Such observation suggests AEG a promising approach, as it is typically more difficult to model the valence perception from audio signals \cite{yang11book}.

\begin{table}[!t]
\centering
\small
\caption{Performance evaluation on general MER ($\downarrow$ stands for smaller-better and $\uparrow$ larger-better).}{
\begin{tabular}{|l|c|c|c|c|}
\hline
 Method & AKL $\downarrow$ & AED $\downarrow$ & ~$R^2$ Valence $\uparrow$~ & ~$R^2$ Arousal $\uparrow$~ \\
\hline\hline
Base-rate       & 1.2228  &	0.4052   &	--0.0009   &	0.0000    \\
SVR-RBF         & 0.7124	&   0.2895   &	 0.1409   &	0.6613    \\
AEG ($K=128$)   & \textbf{0.7049}  &	0.2890   &	\textbf{0.1601}  &	0.6554    \\
AEG ($K=256$)~~ & ~~0.7078~~	&  ~~\textbf{0.2869}~~   &	0.1579  &	\textbf{0.6686}  \\
\hline
\end{tabular}}
\label{tab: mer_summary}
\end{table}

\begin{figure}[!t]
\centering
\subfigure[AKL, smaller-better.]{\includegraphics[width=.44\textwidth]{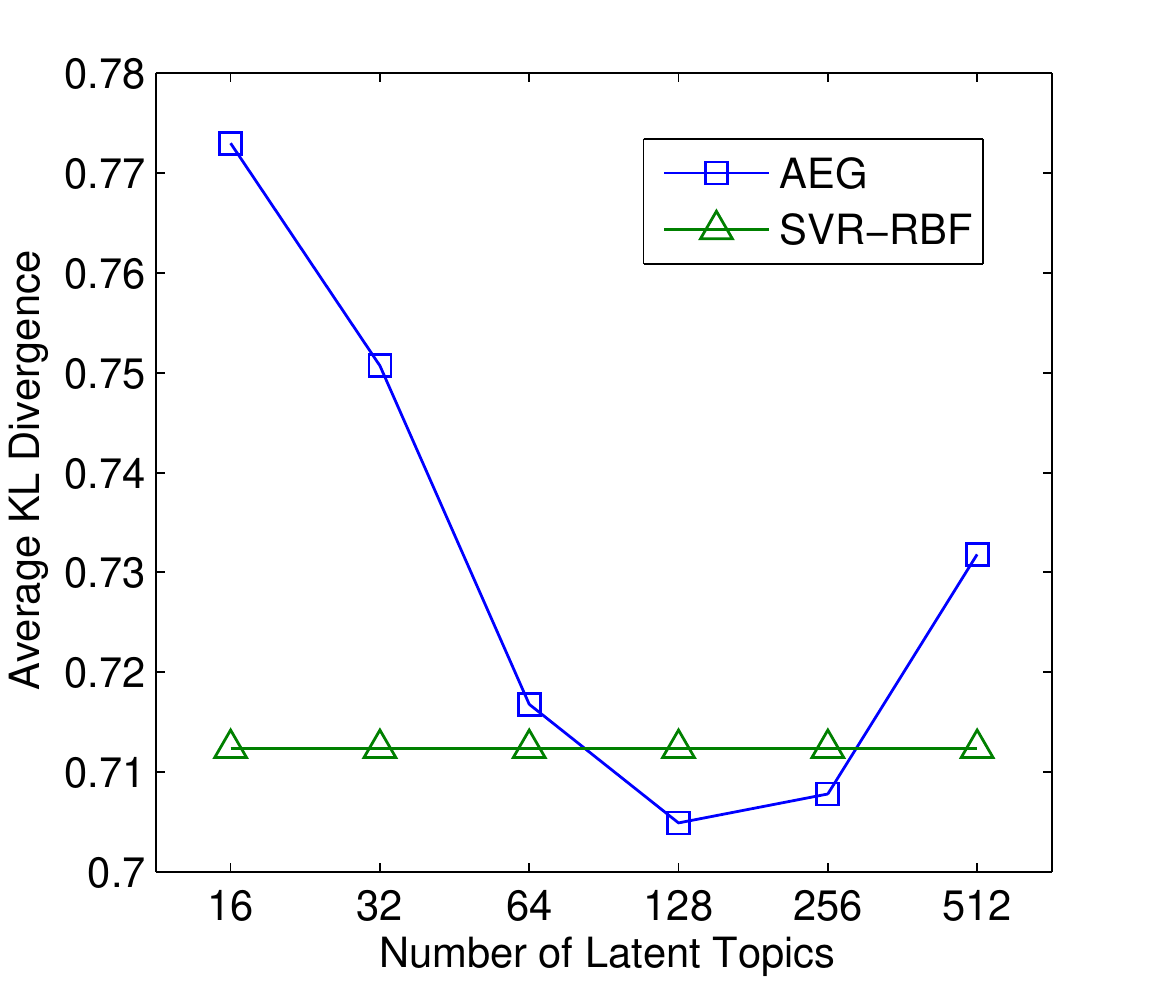}}
\subfigure[AED, smaller-better.]{\includegraphics[width=.44\textwidth]{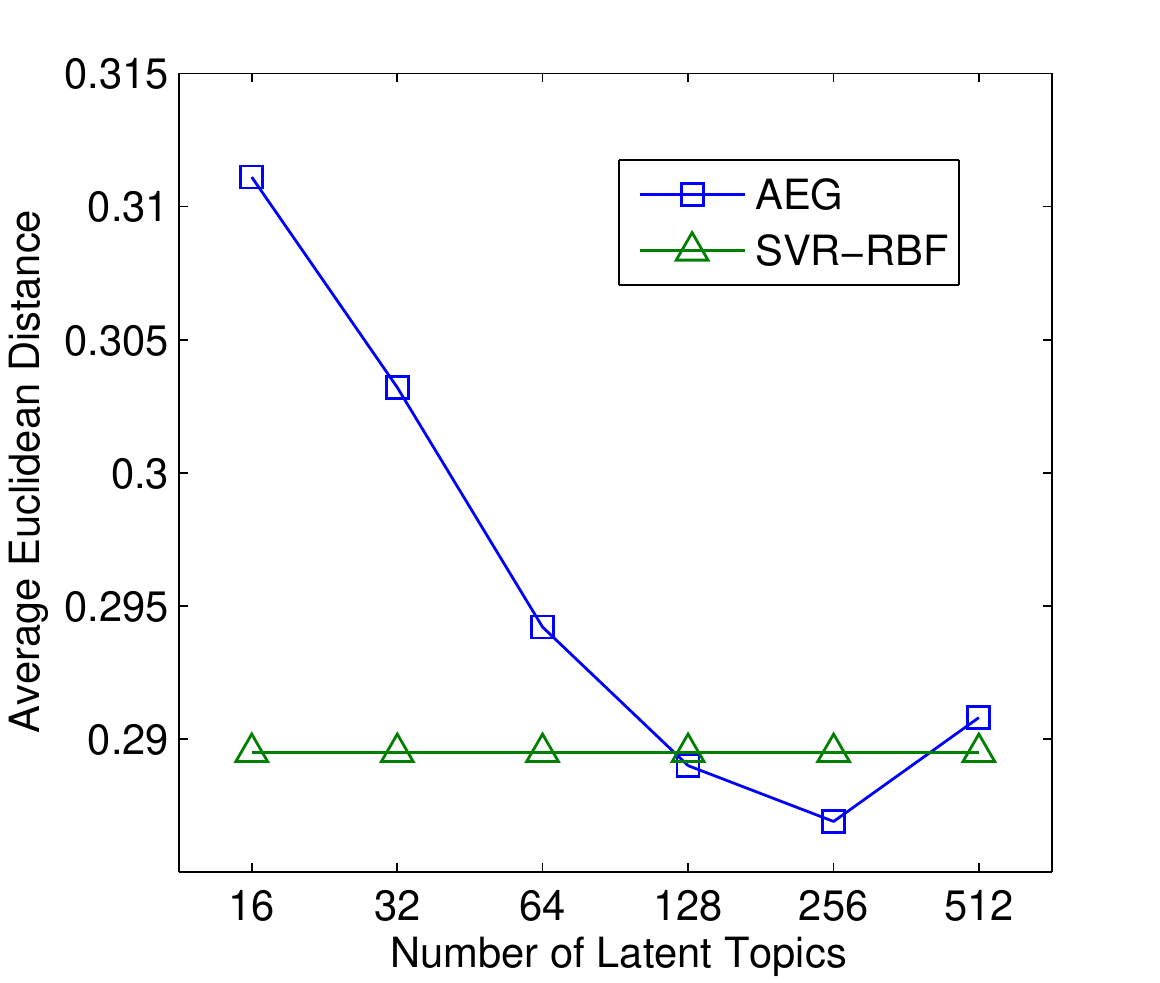}}
\subfigure[$R^2$ of valence, larger-better.]{\includegraphics[width=.42\textwidth]{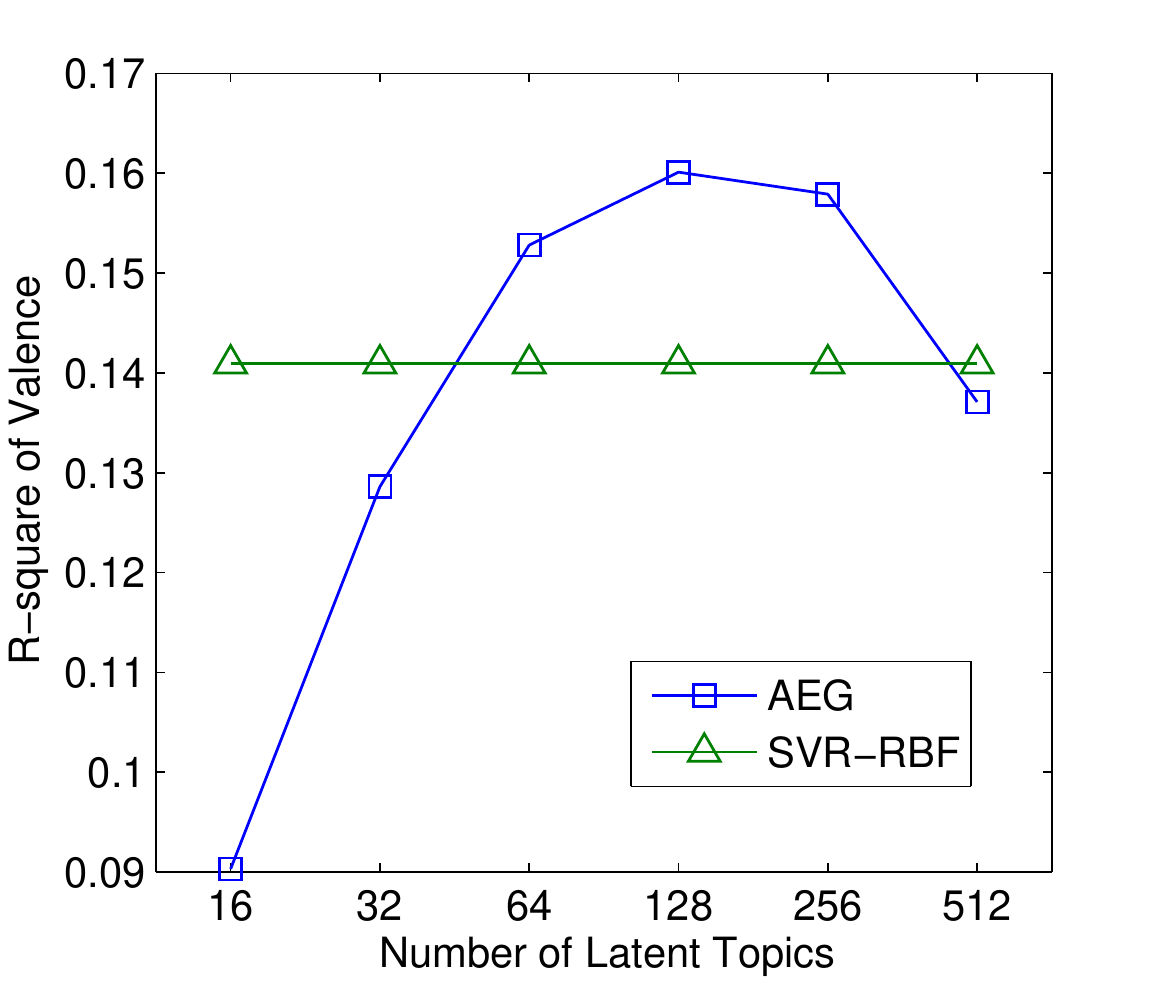}}
\subfigure[$R^2$ of arousal, larger-better.]{\includegraphics[width=.42\textwidth]{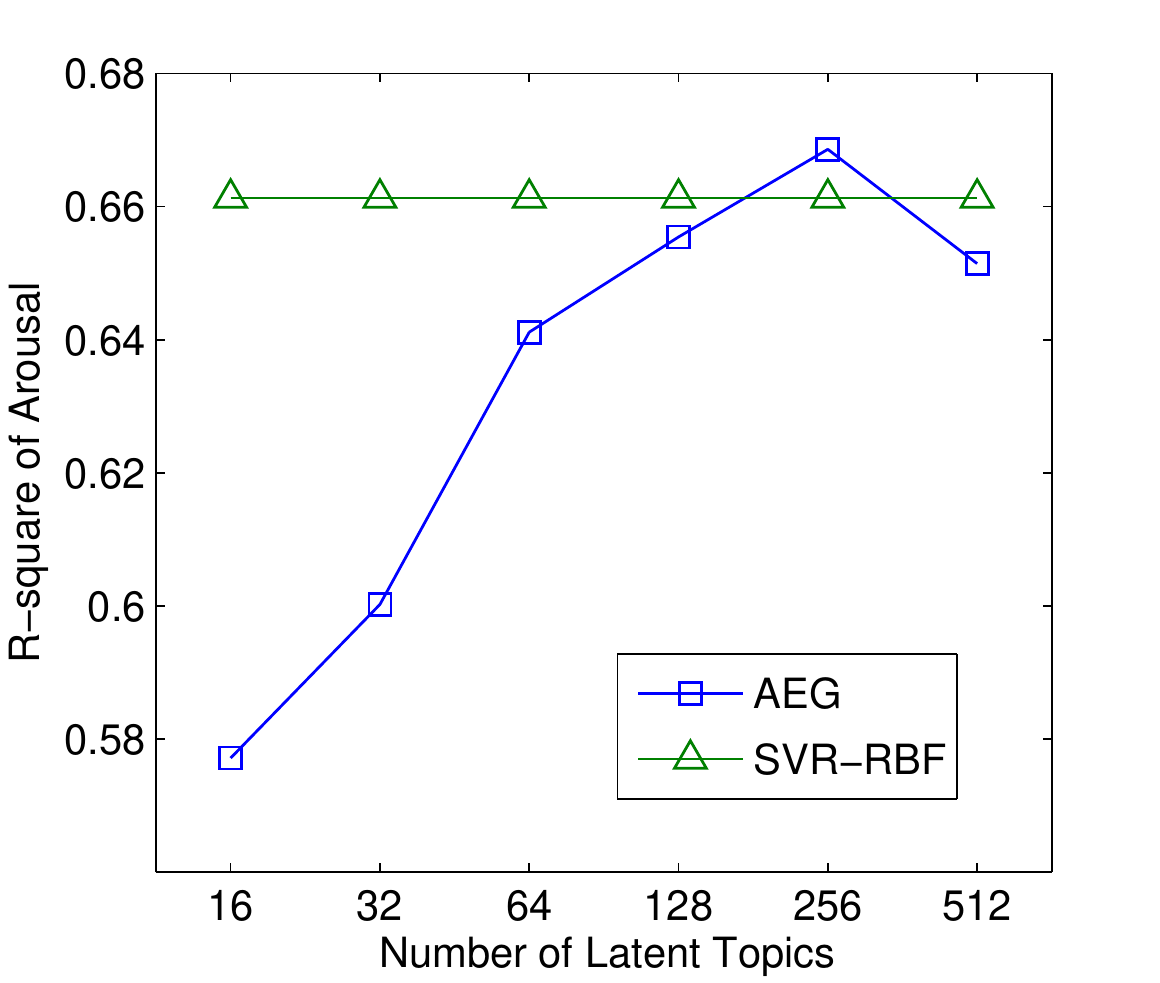}}
\caption{Performance evaluation on general MER, using different numbers of latent topics in AEG.}
\label{fig: EMO_RES1}
\end{figure}

Figure \ref{fig: EMO_RES1} presents the result of AEG when we vary the value of $K$ (i.e. the number of latent topics). It can be seen that the performance of AEG improves as a function of $K$ when $K$ is smaller than 256, but starts to decrease when $K$ is sufficient larger.
The best result is obtained when $K$ is set to 128 or 256.
As the parameters of SVR-RBF has also been optimized, this result shows that, if the optimal case of AEG is not attained (e.g., $K$ = 64 or 512), AEG is still on par with the state-of-the-art SVR approach to general MER.

\subsection{Evaluation on Personalized MER}

\subsubsection{Evaluation Setup}

The trade-off between the number of personal annotations (feedbacks) and the performance of personalization is important for personalized MER. On one hand, we hope to have more personal annotations to more accurately model the emotion perception of a particular user. On the other hand, we want to restrict the number of personal annotations so as to relieve the burden on the user. To reflect this, evaluation on the performance of personalized MER is conducted by fixing the test set for each user, but varying the number of available emotion annotations from the particular user to test how the performance improves as personal data amasses.

We consider 41 users who have annotated more than 150 clips in this evaluation. We use the data of 6 of them for parameter tuning, and the data of the remaining 35 in the evaluation and report the average result for these 35 test users.
One hundred annotations of each test user are randomly selected as the personalized training set for personalization for the user.
Once the model is created, another 50 clips annotated by the same user are randomly selected.
Specifically, for each test user, a general MER model is trained with 600 clips randomly selected from the original AMG1608, excluding those annotated by the test user under consideration and those self-inconsistent annotations.
Then, the general model is incrementally personalized five times using different numbers of clips selected from the personalized training set.
We use 10, 20, 30, 40, and 50 clips iteratively, with the preceding clips being a subset of the current ones each time.
The process is repeated 10 times for each user.



We use the following evaluation metrics here: the AED, the $R^2$, and the average likelihood (ALH) of generating the ground-truth annotation (a single VA point) $\mathbf{e}_\star$ of the test user using the predicted affective Gaussian, i.e. $p(\mathbf{e}_\star \mid  \bm{\hat\mu}_\star, \bm{\hat\Sigma}_\star )$. Larger ALH corresponds to better accuracy.
We do not report KL divergence here because each clip in the dataset is annotated by a user at most once, which does not constitute a probability distribution.

\subsubsection{Result}

We compare the MAP-based personalization method of AEG with the two-stage personalization method of SVR proposed in \cite{yang07hcm}. In the two-stage SVR method, the first stage creates a general SVR model for general emotion prediction, whereas the second stage creates a personalized SVR that is trained solely on a user's annotations. The final prediction is obtained by linearly combining the predictions from the general SVR and the personalized SVR with weights 0.7 and 0.3, respectively. The weights are derived empirically according to our pilot study. As for AEG, we only update the mean parameters with $\beta^\text{m} = 0.01$, because our pilot study shows that updating the covariance empirically does not lead to better performance. This observation is also in line with the findings in speaker adaptation \cite{reynolds2000dsp}. We train the background model with AEG Uniform for simplicity.

\begin{figure}[!t]
\centering
\subfigure[ALH, larger-better.]{\includegraphics[width=.45\textwidth]{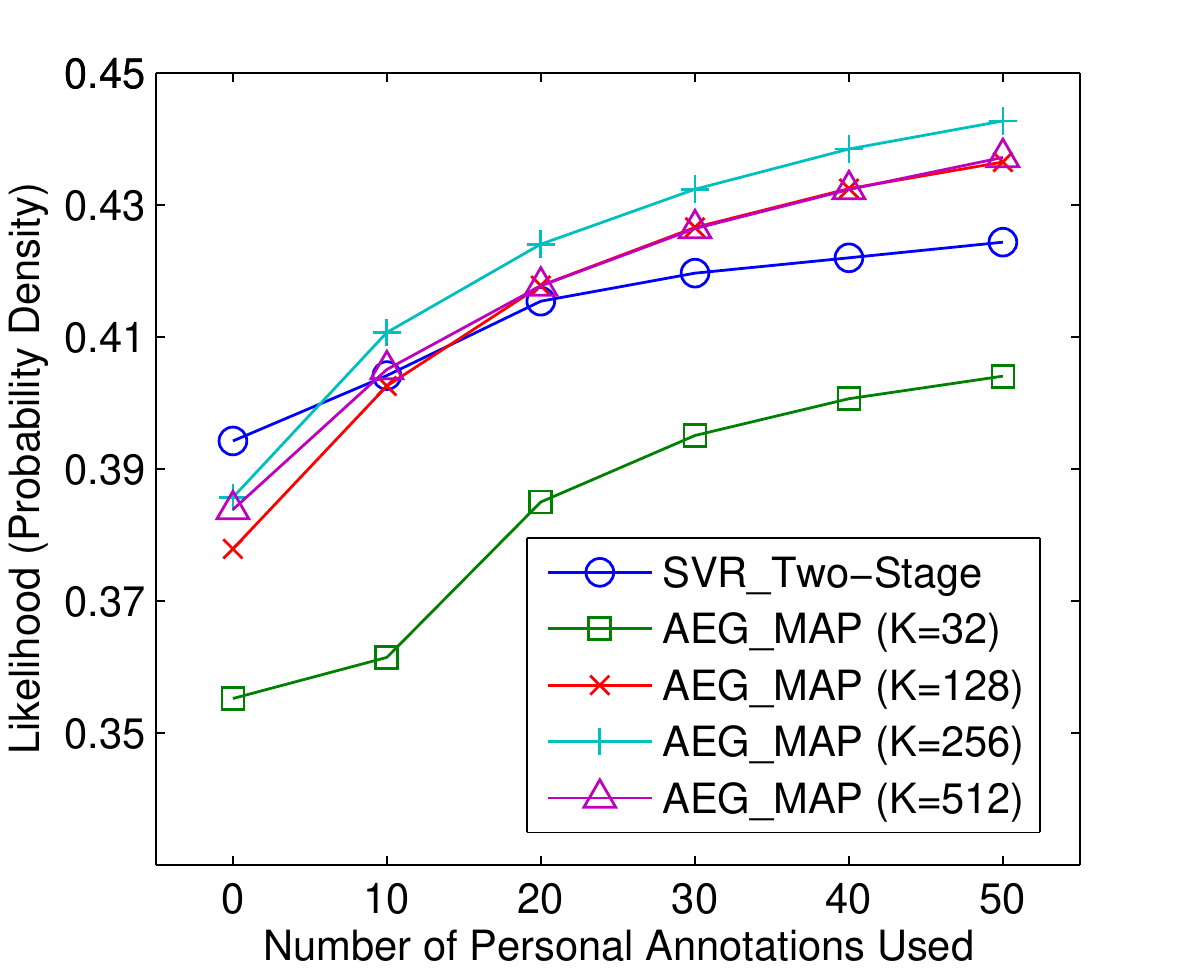}}
\subfigure[AED, smaller-better.]{\includegraphics[width=.45\textwidth]{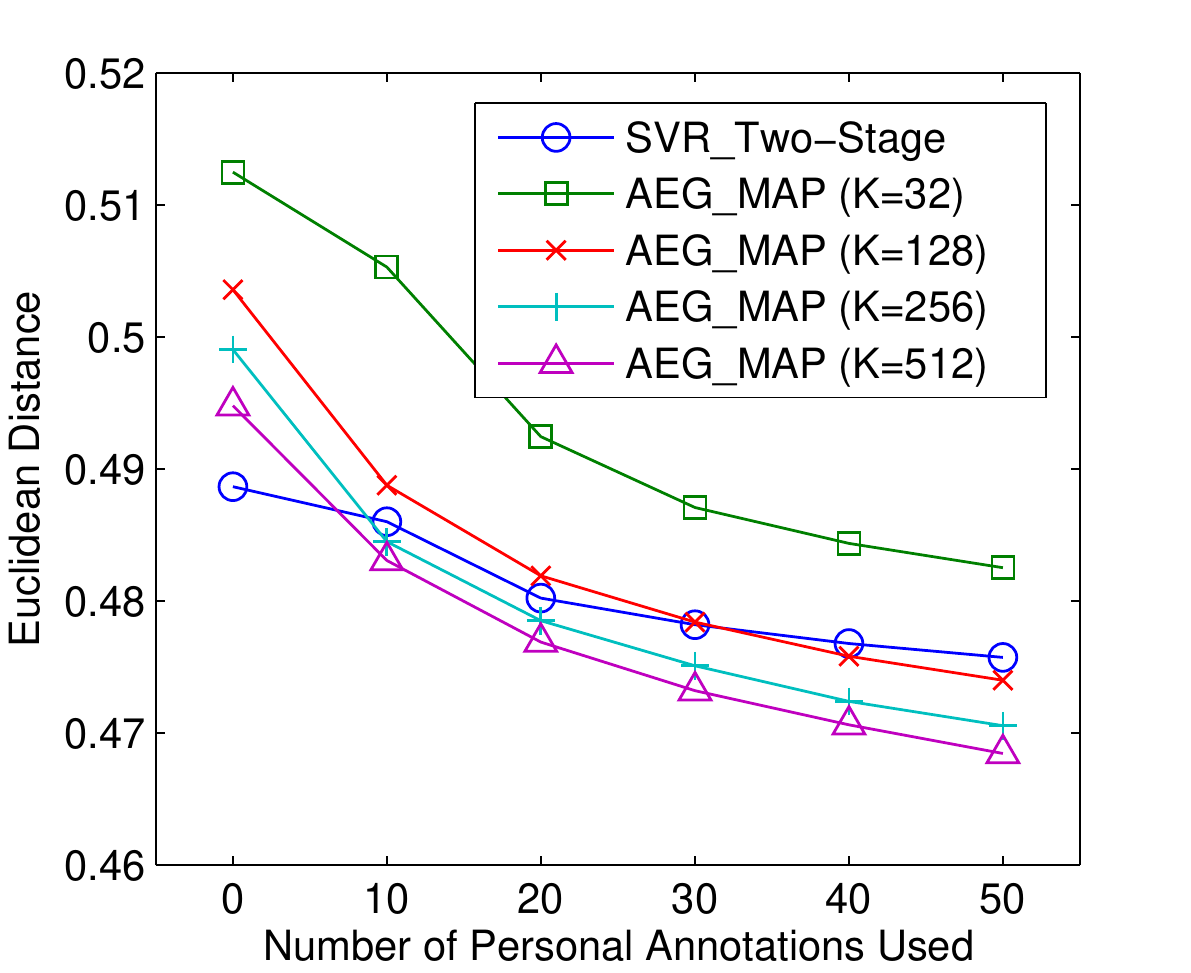}}
\subfigure[$R^2$ of valence, larger-better.]{\includegraphics[width=.45\textwidth]{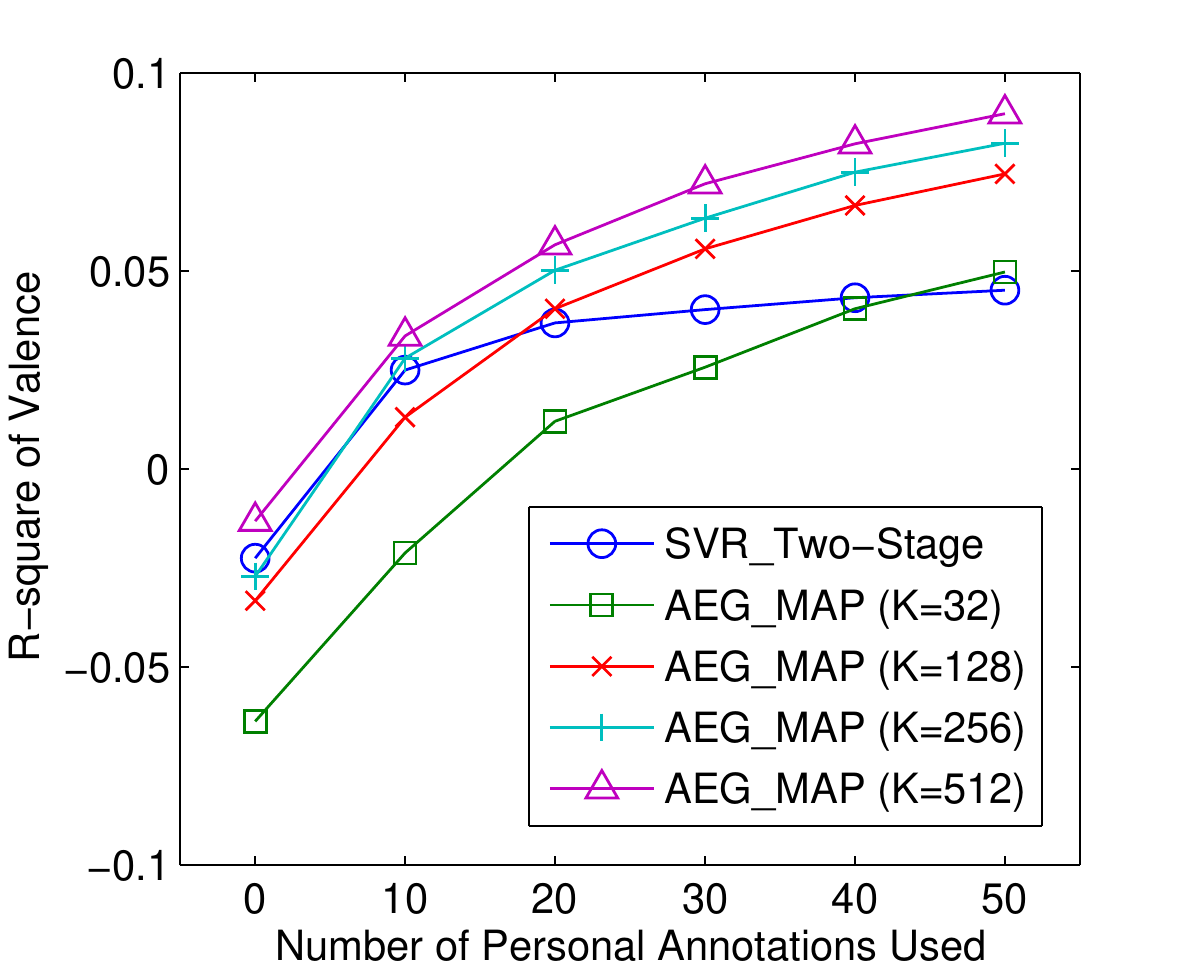}}
\subfigure[$R^2$ of arousal, larger-better.]{\includegraphics[width=.45\textwidth]{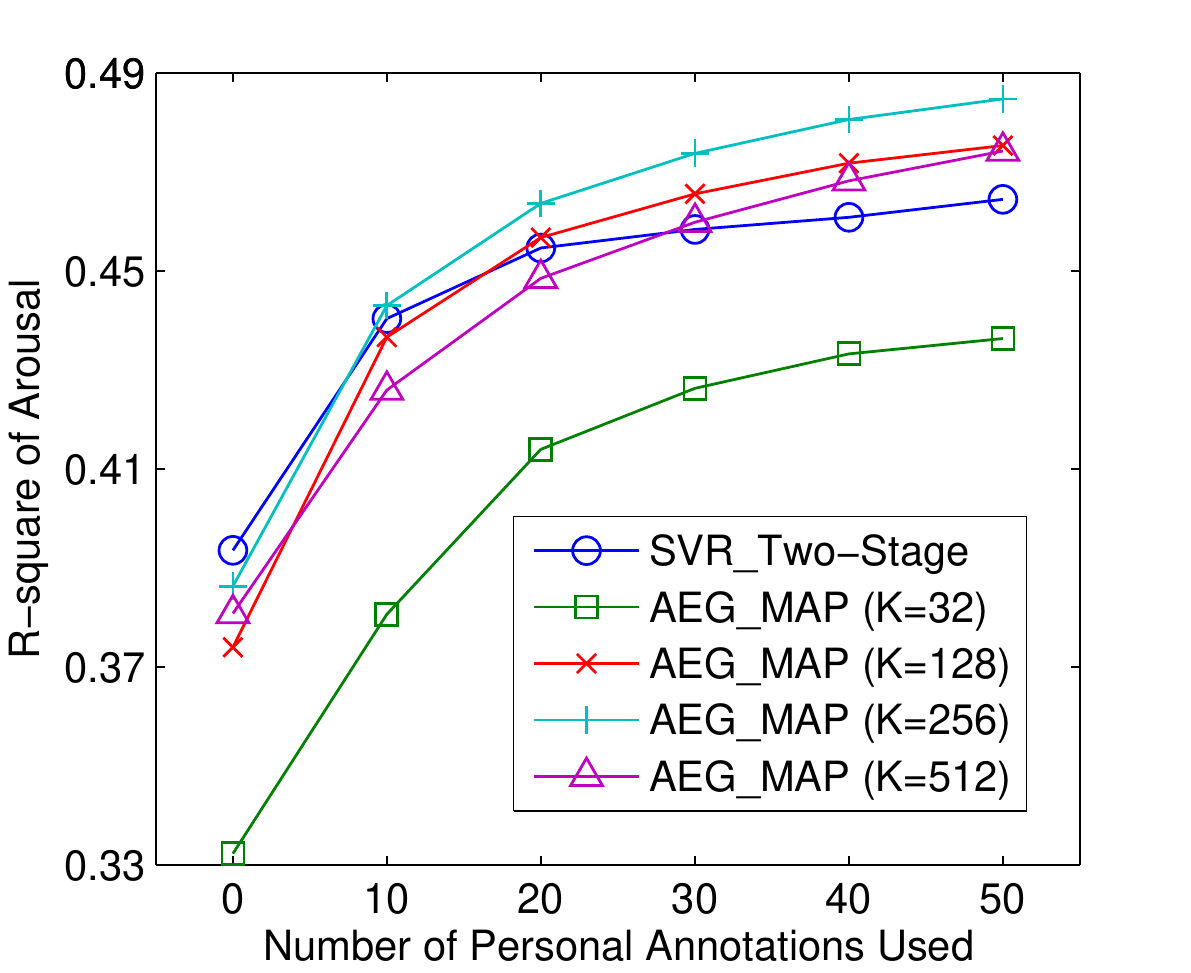}}
\caption{Performance evaluation on personalized MER, with varying numbers of personal data.}
\label{fig: EMO_RES2}
\end{figure}

Figure \ref{fig: EMO_RES2} compares the result of different personalized MER methods, when we vary the number of available personal annotations. The starting point of each curve is the result given by the general MER model trained on partial users of the AMG1608 dataset.
We can see that the result of the general model is inferior to those reported in Figure \ref{fig: EMO_RES1}, showing that a general MER model is less effective when it is used to predict the emotion perception of individual users, compared to the case of predicting the \emph{average} emotion perception of users. We can also see that the result of the considered personalized methods generally grows as the number of personal annotations increases. When the value of $K$ is sufficiently large, AEG-based personalization methods can outperform the SVR method. Moreover, while the result of SVR starts to saturate when the number of personal annotations is larger than 20, AEG has the potential of keeping on improving the performance by exploiting more personal annotations. We also note that there is no significant performance difference for AEG when $K$ is large enough (e.g. $\geq 128$).

Although our evaluation shows that personalization methods can improve the result of personalized emotion prediction, the low values in the $R^2$ statistics for valence and arousal still show that the problem is fairly challenging.  Future work is still needed to improve either the quality of the emotion annotation data or the feature extraction or machine learning algorithms for modeling emotion perception.

\section{Emotion-based Music Retrieval}
\label{sec: EMO_MR}

\subsection{The VA-oriented Query Interface}

The VA space offers a ready canvas for music retrieval through the specification of a point in the emotion space \cite{mremo}. Users can retrieve music pieces of certain emotions without specifying the titles.
Users can also draw a trajectory to indicate the desired emotion changes across a list of songs (e.g. from angry to tender).

\begin{figure}[!t]
\centering
\subfigure[Point-based]{
\includegraphics[height=.43\textwidth]{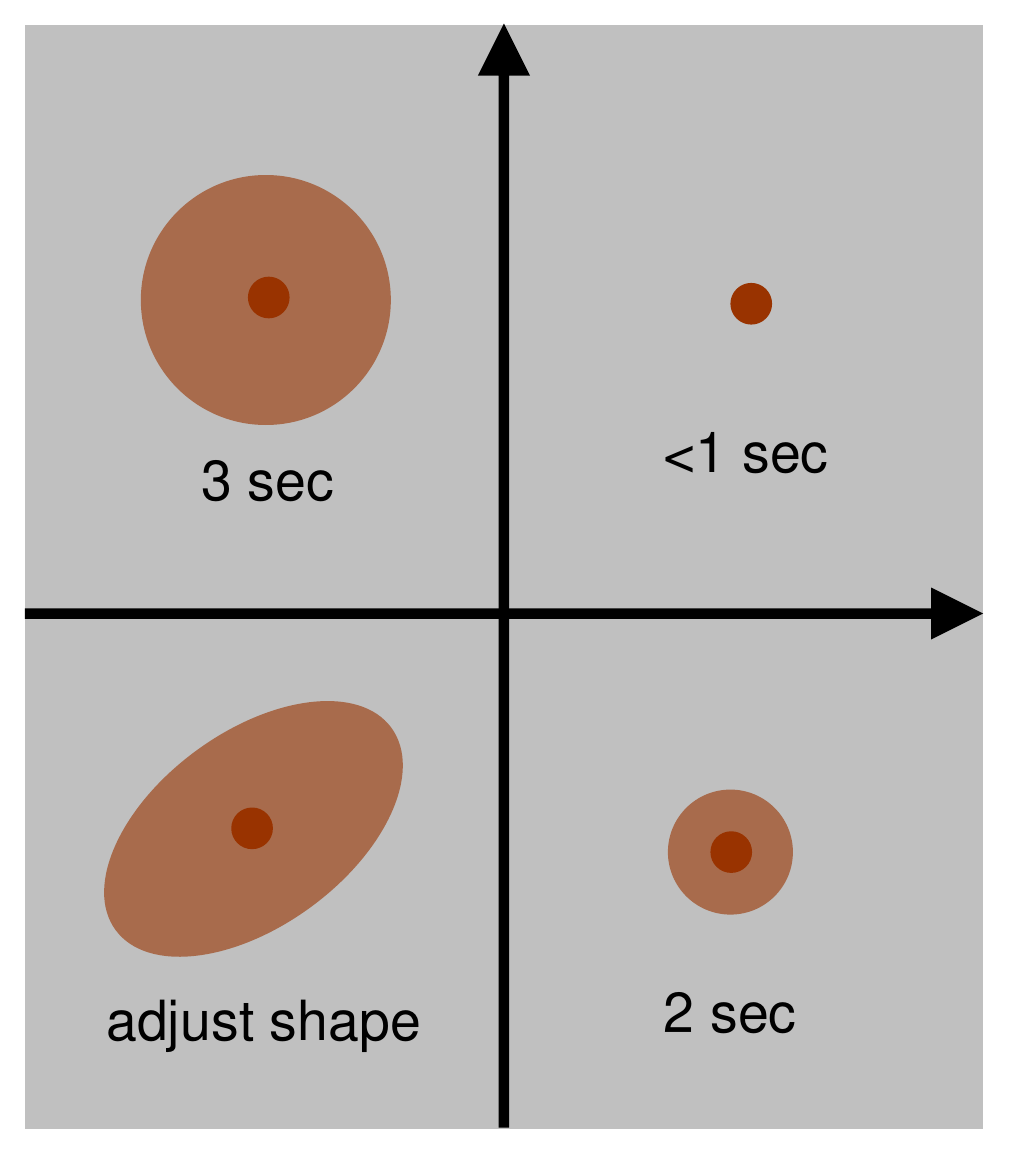}
\label{fig: EMO_APPA}
}~~~~~~~
\subfigure[Trajectory-based]{
\includegraphics[height=.43\textwidth]{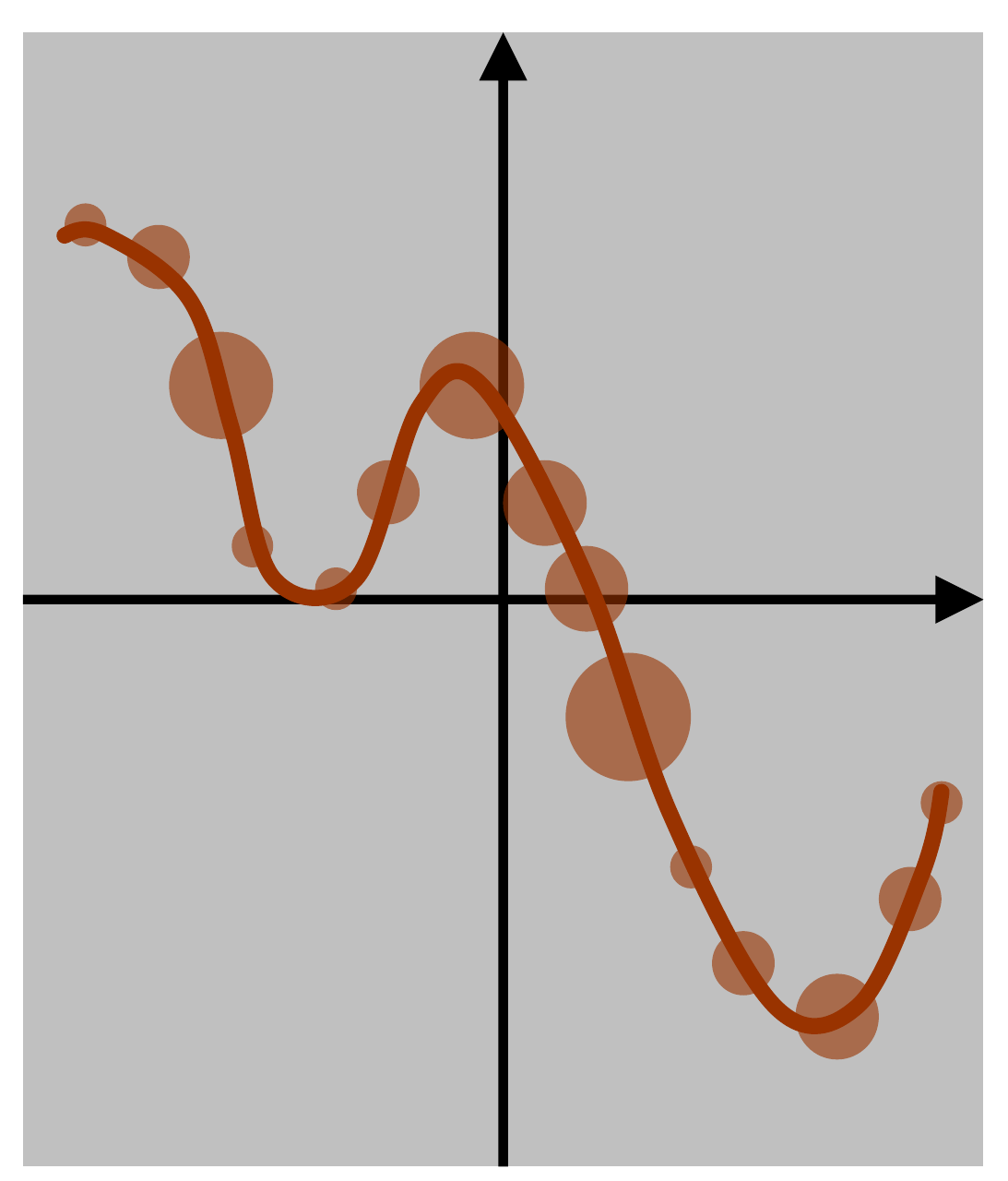}
\label{fig: EMO_APPB}
}
\caption{The stress-sensitive user interface for emotion-based music retrieval. Users can (a) specify a point or (b) draw a trajectory, while specifying the variance with different levels of duration.}
\label{fig: EMO_APP}
\end{figure}

In addition to the above point-based query, one can also issue a Gaussian-based query to an AEG-based retrieval system. As Figure \ref{fig: EMO_APP} shows, users can specify the desired variances (or the confidence level at the center point) of emotion by pressing a point in the VA space with different levels of duration or strength.
The variance of the Gaussian gets smaller as one increases the duration or strength of pressing, as Figure \ref{fig: EMO_APP}~(a) shows. Larger variances indicate less specific emotion around the center point. After specifying the size of a circular variance shape, one can even pinch fingers to adjust the variance shape.
For a trajectory-based query input, similarly, the corresponding variances are determined according to the dynamic speed when drawing the trajectory, as Figure \ref{fig: EMO_APP}~(b) shows. Fast speed corresponds to a less specific query and the system will return pieces whose variances of emotion are larger. If songs with more specific emotions are desirable, one can slow down the speed when drawing the trajectory.
The queries inputted by such a \emph{stress-sensitive interface} can be handled by AEG for emotion-based music retrieval.

\subsection{Overview of the Emotion-based Music Retrieval System}

\begin{figure}[!t]
\centering
\includegraphics[width=.7\columnwidth]{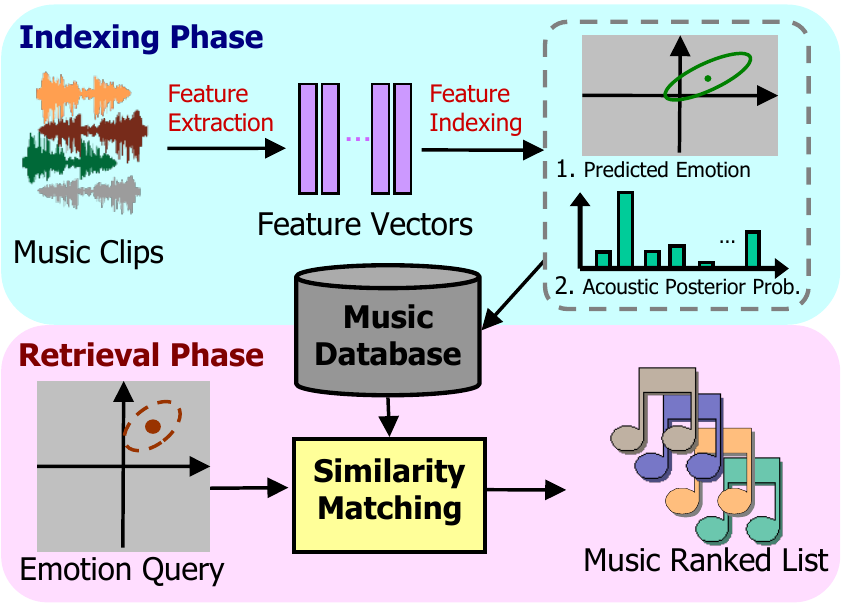}
\caption{The diagram of the content-based music retrieval system using an emotion query.}
\label{fig: EMO_SYS2}
\end{figure}

As Figure \ref{fig: EMO_SYS2} shows, the content-based retrieval system can be divided into two phases. In the \emph{feature indexing} phase, we index each music clip in an un-labeled music database by one of the following two approaches: The \emph{emotion prediction} approach indexes a clip with the \emph{predicted emotion distribution} (an affective GMM or a single 2-D Gaussian) given by MER, whereas the \emph{folding-in} approach indexes a clip with the \emph{topic posterior} (a $K$-dimensional vector).
In the later \emph{music retrieval} phase, given an arbitrary emotion-oriented query the system returns a list of music clips ranked according to one of the following two approaches:
\emph{likelihood/distance-based matching} and \emph{pseudo song-based matching}. These two ranking approaches correspond to one of the two indexing approaches, respectively, as summarized in Table \ref{tab:retreival_method}. We present the details of the two approaches in the following subsections.

\begin{table}[!t]
\centering
\footnotesize
\caption{The two approaches of the emotion-based music retrieval system}{
\begin{tabular}{|p{2cm}|p{2.7cm}|p{3cm}|p{4cm}|}
\hline
Approach & Indexing phase & Indexed type & Matching phase  \\
\hline\hline
\textbf{Emotion ~~~Prediction}  & full procedure of MER by AEG & an affective GMM (Eq. \ref{eq: predict_form}) or a 2-dim Gaussian $\{\bm{\hat \mu}, \bm{\hat\Sigma}\}$
& likelihood (for point query) ~~~or distance (for Gaussian query) \\
\hline
\multirow{2}{*}{\textbf{Folding-In}} & compute only the topic posterior & \multirow{2}{*}{$K$-dim vector $\bm{\hat\theta}$}
& cosine similarity of pseudo song ~~~ ($K$-dim vector $\bm{\lambda}$) \\
\hline
\end{tabular}}
\label{tab:retreival_method}
\end{table}

\subsection{The Emotion Prediction-based Approach}
\label{sec: EMO_MR_EmotionPrediction}

This approach indexes each clip as a single, representative Gaussian distribution or an affective GMM in the offline MER procedure. The query is then used to compare with the predicted emotion distribution of each clip in the database. The system ranks all the clips based on the likelihoods or distances in response to the query. Clips with larger likelihood or smaller distance should be placed in the higher order.

Given a point query ${\mathbf{\tilde e}}$, the corresponding likelihood of the indexed emotion distribution of a clip $\hat{\bm{\theta}}_i$ is generated by a single Gaussian $p(\mathbf{\tilde e} \mid \hat{\bm{\mu}}_i, \bm{\hat\Sigma}_i)$ or an affective GMM $p(\mathbf{\tilde e} \mid \hat{\bm{\theta}}_i)$ (cf. Eq. \ref{eq: predict_form}), where $\{\hat{\bm{\mu}}_i, \bm{\hat\Sigma}_i\}$ is the predicted parameters of the representation Gaussian for $\hat{\bm{\theta}}_i$, and $\hat\theta_{i,k}$ is the $k$-th component of $\bm{\hat\theta}_i$. Note that here we use the topic posterior vector to represent a clip in the database.

When it comes to a Gaussian-based query $\tilde{G}=G(\bm{\tilde\mu}, \bm{\tilde\Sigma})$, the approach generates the ranking scores based on the KL2 divergence. In the case of indexing with a single Gaussian, we use Eq. \ref{eq: kl_two} to compute $D_\text{KL2} \big( \tilde{G}, G(\bm{\hat\mu}_i, \bm{\hat\Sigma}_i) \big)$ between the query and a clip. On the other hand, in the case of indexing with an affective GMM, we compute the weighted KL2 divergence by
\begin{equation}
D_\text{KL2}\big( \tilde{G}, p(\mathbf{e} \mid \bm{\hat\theta}_i ) \big) =
\sum_{k=1}^K \hat\theta_{i,k}  D_\text{KL2} \big( \tilde{G}, G_k( \bm{\mu}_k, \bm{\Sigma}_k) \big)\,.
\label{eq: wKL}
\end{equation}

\subsection{The Folding-In-based Approach}
\label{sec: EMO_MR_FoldingIn}

\begin{figure}[!t]
\centering
\includegraphics[width=.7\textwidth]{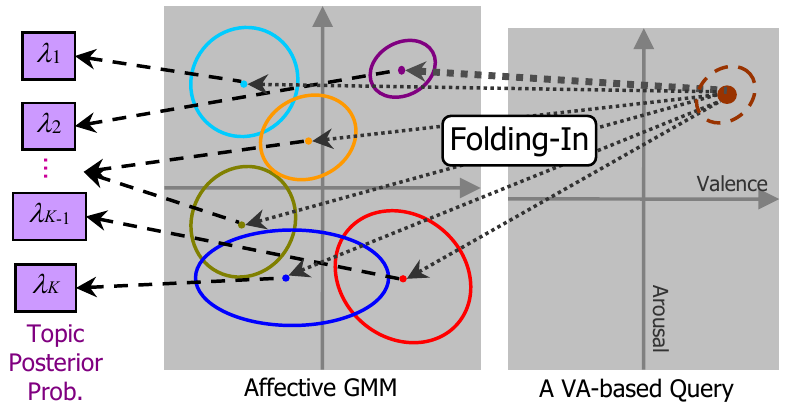}
\caption{Illustration of the Folding-In process of emotion-based music retrieval by AEG.}
\label{fig: EMO_SYS3}
\end{figure}

As Figure \ref{fig: EMO_SYS3} shows, this approach estimates the probability distribution $\bm{\lambda} = \{\lambda_k\}_{k=1}^K$, subject to $\sum_k{\lambda_k}=1$, for an input VA-oriented query in an online manner. Each estimated $\lambda_k$ corresponds to the relevance of a query to the $k$-th latent topic $z_k$, so we can treat the distribution of $\bm{\lambda}$ as the topic posterior of the query and call it a \emph{pseudo song}. In the case of Figure \ref{fig: EMO_SYS3}, for example, we show a query that is very likely to be represented by the 2-nd affective Gaussian component. The folding-in process is likely to assign a dominative weight $\lambda_2 =1$ for $z_2$, and $\lambda_h=0$, $\forall h\neq2$. This implies that the query is highly related to the song whose topic posterior is dominated by $\theta_2$. Therefore, the pseudo song can be used to match with the topic posterior vector $\bm{\hat\theta}_i$ of each clip in the database.


Given a point query ${\mathbf{\tilde e}}$, we start the folding-in process by first generating the pseudo song via maximizing the query likelihood of the $\bm{\lambda}$-weighted affective GMM with respective to $\bm{\lambda}$. By taking the logarithm of Eq. \ref{eq: predict_form}, we obtain the following objective function,
\begin{equation}
\underset{\bm{\lambda}} {\max} ~~ \log \sum_{k=1}^K \lambda_k ~ G_k( \mathbf{\tilde e} \mid \bm{\mu}_k, \bm{\Sigma}_k) \,,
\label{eq: lambda}
\end{equation}
where $\lambda_k$ is the $k$-th component of the vector $\bm{\lambda}$.
In some sense, a good $\bm{\lambda}$ will make the corresponding $\bm{\lambda}$-weighted affective GMM well generate the query ${\mathbf{\tilde e}}$. The problem in Eq. \ref{eq: lambda} can be solved by the EM algorithm. In the E-step, the posterior probability of $z_k$ is computed by
\begin{equation}
p(z_k \mid \mathbf{\tilde e}) = \frac{ \lambda_k G_k( \mathbf{\tilde e} \mid \bm{\mu}_k, \bm{\Sigma}_k ) } {\sum\nolimits_{h=1}^K \lambda_h G_h( \mathbf{\tilde e} \mid \bm{\mu}_h, \bm{\Sigma}_h ) }\,.
\label{eq: p(z|e)}
\end{equation}
In the M-step, we then only update $\lambda_k$ by
\begin{equation}
\lambda'_k \leftarrow p(z_k \mid \mathbf{\tilde e}) \,.
\label{eq: lambda_upt1}
\end{equation}

As for a Gaussian-based query $\tilde{G}$, we fold in the query into the learned affective GMM to estimate a pseudo song as well. This time, we maximize the following log-likelihood function,
\begin{equation}
\underset{\bm{\lambda}} {\max} ~~  \log \sum_{k=1}^K \lambda_k ~p( \tilde{G} \mid  G_k) \,,
\label{eq: NMM}
\end{equation}
where $p(\tilde{G} \mid G_k )$ is the likelihood function based on KL2 (cf. Eq. \ref{eq: kl_two}):
\begin{equation}
p(\tilde{G} \mid G_k ) = \exp \big( -D_\text{KL2}( \tilde{G}, G_k) \big) \,.
\label{eq: KL2_likelihood}
\end{equation}
Again, Eq. \ref{eq: NMM} can be solved by the EM algorithm, with the following update,
\begin{equation}
\lambda'_k \leftarrow p(z_k \mid \tilde{G}) =
\frac{ \lambda_k p( \tilde{G} \mid G_k)} {\sum\nolimits_{h=1}^K \lambda_h p( \tilde G \mid G_h)} \,.
\label{eq: p(z|N)}
\end{equation}

The EM processes for both point- and Gaussian-based queries stop early after few iterations (e.g. 3), because the pseudo song estimation is sensitive to over-fitting. Several initialization settings can be used, such as a random, uniform, or prior distribution. Considering the stability and the reproducibility of the experimental result, we opt for using a uniform distribution for initialization. Note that random initialization may introduce discrepant results among different trials even with identical experimental settings, whereas initializing with a prior distribution may render biased results in favor of songs that predominates the training data \cite{wang12acmmm}.
Finally, the retrieval system ranks all the clips in descending order of the following cosine similarities in response to the pseudo song:
\begin{equation}
\Phi(\bm{\lambda}, \bm{\theta}_i) = \frac{\bm{\lambda}^T \bm{\theta}_i} {\|\bm{\lambda}\| \|\bm{\theta}_i\| }\,.
\label{eq: cossim}
\end{equation}

\subsection{Discussion}

The Emotion Prediction approach is straightforward, as the purpose of MER is to automatically index unseen music pieces in the database. In contrast, the Folding-In approach goes one step further to embed a VA-based query into the space of music clips. Although the folding-in process requires an additional step of estimating the pseudo song, it is in fact more flexible. In a personalized music retrieval context, for example, a personalized affective GMM can readily produce a personalized pseudo song for comparing with the original topic posterior vectors of all the pieces in the database, without the need to predict the emotion again with the personalized model.

The complexity of the Emotion Prediction approach mainly comes from computing the likelihood of a point query on each music clip's emotion distribution or the KL divergence between the Gaussian query and the emotion distribution of each clip. Therefore, the matching process needs to compute $N$ (the number of clips in the database) times the likelihood or the KL divergence. In the Folding-In approach, the complexity comes from estimating the pseudo song (with the EM algorithm) and computing the cosine similarity between the pseudo song and each clip. EM needs to compute $K\times ITER$ times the likelihood of a component affective Gaussian or the Gaussian KL divergence, where $ITER$ is the number of EM iterations. Then, the matching process computes $N$ times the cosine similarity. Obviously, computing the likelihood on an emotion distribution (i.e. a single Gaussian or a GMM) is computationally more expensive than computing the cosine similarity (as $K$ is usually not large). Therefore, when $N$ is large (e.g. $N \gg K\times ITER$), the Folding-In approach is considered as a more feasible one in practice.

\subsection{Evaluation for Emotion-based Music Retrieval}
\label{sec: EXP-RETRI}

\subsubsection{Evaluation Setup}

\begin{figure}[!t]
\centering
\subfigure[] {\includegraphics[width=.49\textwidth]{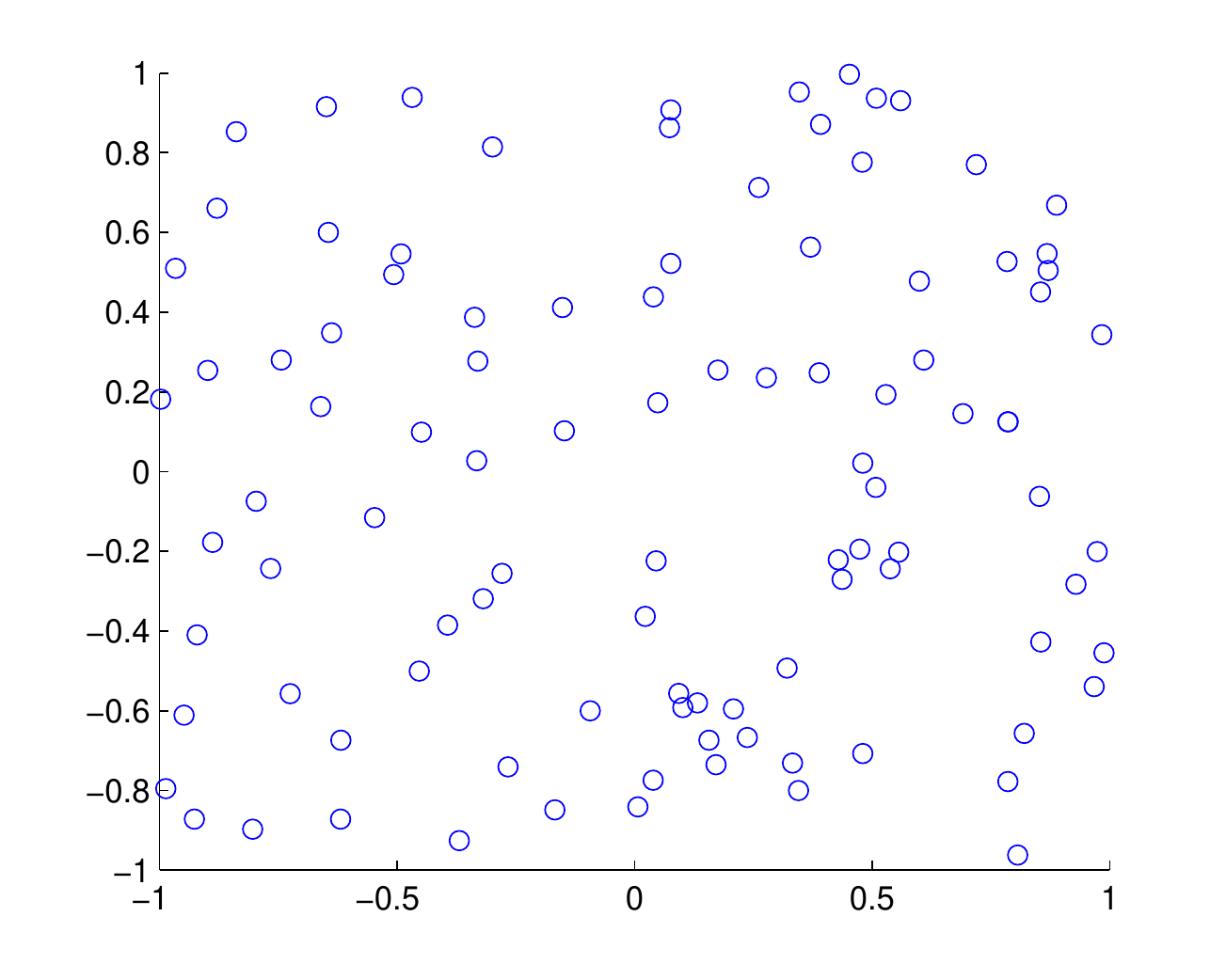}} 
\subfigure[] {\includegraphics[width=.49\textwidth]{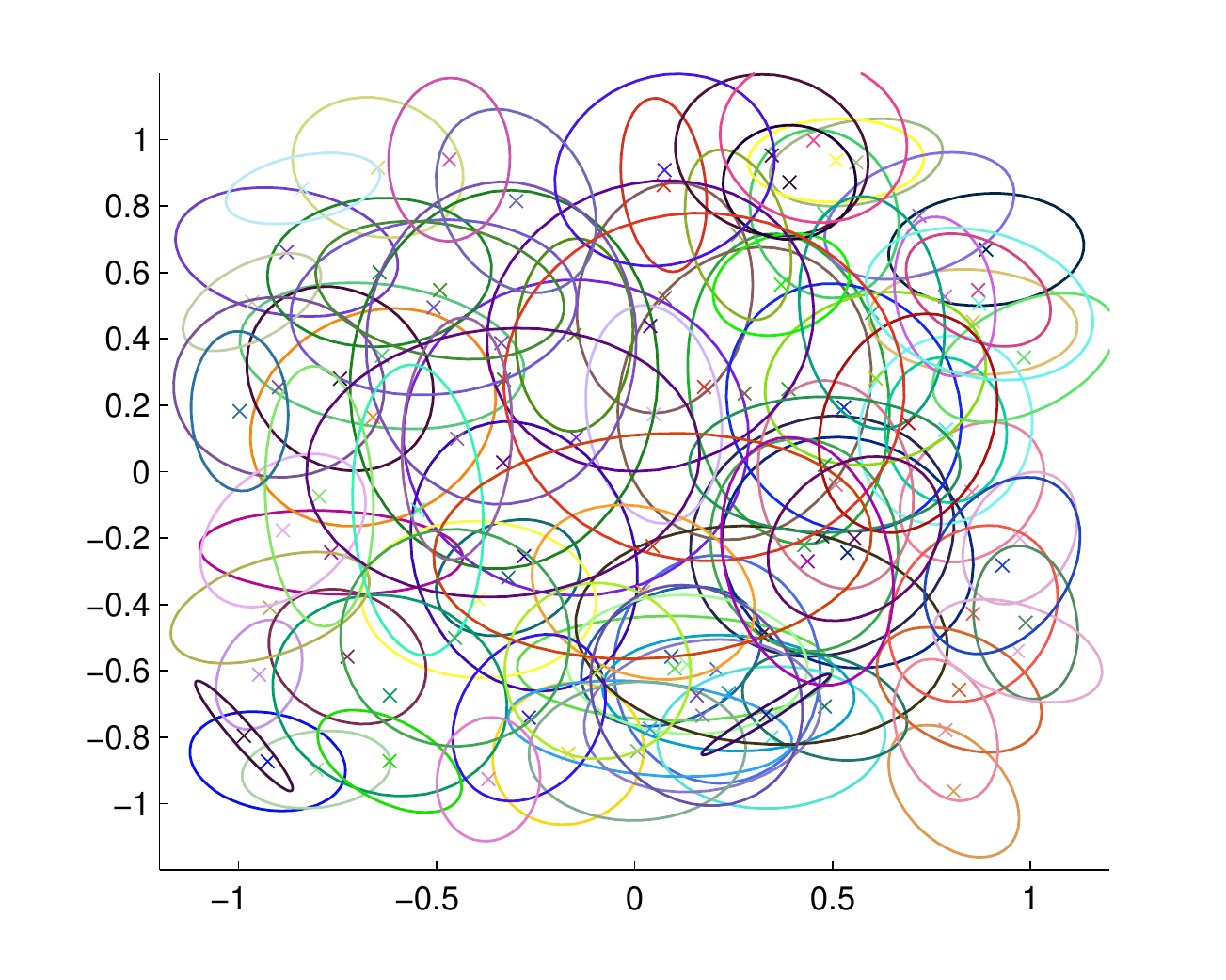}}
\caption{Test queries used in evaluating emotion-based music retrieval: (a) 100 points generated uniformly in between [--1,~1]. (b) 100 Gaussians generated based on the previous 100 points. 
}
\label{fig: test_query}
\end{figure}

The AMG1608 dataset is again adopted in this music retrieval evaluation. We consider two emotion-based music retrieval scenarios: query-by-point and query-by-Gaussian. For each scenario, we create a set of synthetic queries and use the learned AEG model to respond to each test query and return a ranked list of music clips from an unlabeled music database.
The generation of the test query set for query-by-point is simple. As Figure \ref{fig: test_query} (a) shows, we uniformly sample 100 2-D query points within $\left[[-1,-1]^T, [1,1]^T\right]$ in the VA space. The test query set for query-by-Gaussian is then based on this set of points. Specifically, we convert a point query to a Gaussian query by associating with the point a 2-by-2 covariance matrix, as Figure \ref{fig: test_query} (b) shows. Motivated by our empirical observation from data, the covariance of a Gaussian query is set in inverse proportion to the distance between the mean of the Gaussian query (determined by the corresponding point query) and the origin of the VA space. That is, if a given point query is far from the origin (with large emotion magnitude), the user may want to retrieve songs with a specific emotion (with smaller covariance ellipse).

The performance is evaluated by aggregating the ground truth \emph{relevance} scores of the retrieved music clips according to the normalized discounted cumulative gain (NDCG), a widely used performance measure for ranking problems \cite{jarvelin02tois}. The NDCG@$P$, which measures the relevance of the top $P$ retrieved clips for a query, is computed by
\begin{equation}
{\text{NDCG}}@P = \frac{1}{Z_P} \left\{ R(1) + \sum_{i=2}^P \frac{R(i)}{\log_2 i} \right\} \,,
\label{eq: ndcg}
\end{equation}
where $R(i)$ is the ground truth relevance score of the rank-$i$ clip, $i=1,\ldots,Q$, where $Q\geq P$ is the number of clips in the music database, and $Z_P$ is the normalization term that ensures the ideal NDCG@$P$ equal 1. Let $\mathcal{N}_i$ (with parameters $\{\mathbf{a}_i, \mathbf{B}_i\}$) denote the ground-truth annotation Gaussian of the rank-$i$ clip. For a point query $\tilde{\mathbf{e}}$, $R(i)$ is obtained by $p(\tilde{\mathbf{e}} \mid \mathbf{a}_i, \mathbf{B}_i)$, the likelihood of the query point. For a Gaussian query $\tilde{\mathcal{N}}$, $R(i)$ is given by $p(\tilde{\mathcal{N}} \mid \mathcal{N}_i)$ defined by Eq. \ref{eq: KL2_likelihood}. From Eq. \ref{eq: ndcg}, we see that if the system ranks the clips in similar order as the descending order obtained on $\{R(i)\}_{i=1}^Q$, we obtain a larger NDCG. We report the average NDCG computed over the test query set.
Note that we do not adopt evaluation metrics, such as the mean average precision (MAP) and the area under the ROC curve (AUC), because currently it is not trivial to set a threshold to binarize $R(i)$.

We perform three-fold cross-validation as that used in evaluating general MER. In each round, the test fold (with 536 clips) serves as the unlabeled music database. 


\subsubsection{Result}

We implement a Random approach to reflect the lower bound performance by using a random permutation for each test query, without taking into consideration any ranking approach.
We further implement an \emph{Ensemble} approach that averages the rankings of a test query given by Emotion Prediction and Folding-In. Specifically, both approaches assign an ordinal number to a clip according to their respective rankings. Then, we average the two ordinal numbers of a clip as a new score, and re-rank all the clips in ascending order of their new scores.

\begin{figure}[!t]
\centering
\subfigure[Point-based query, larger-better.]{\includegraphics[width=.45\columnwidth]{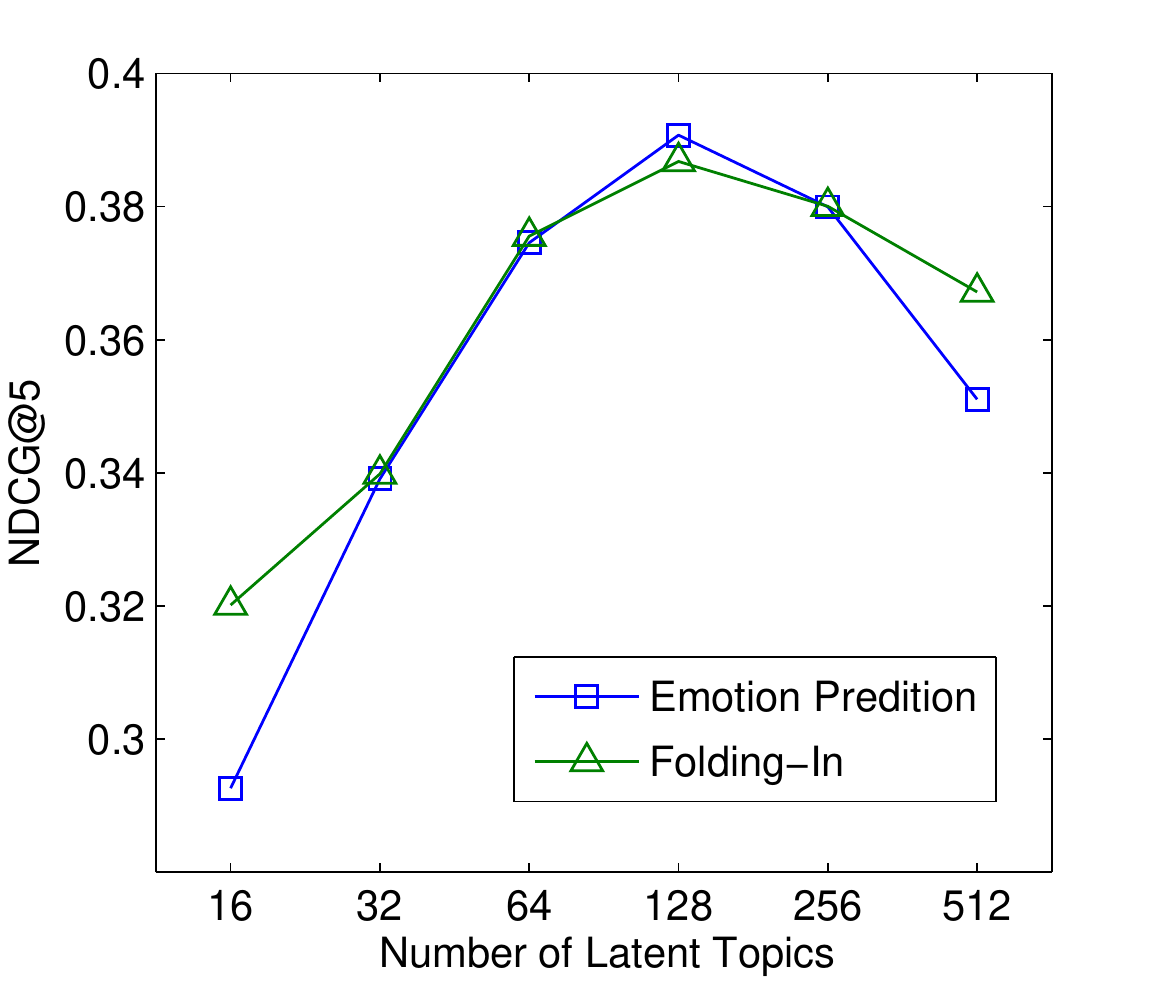}}
\subfigure[Gaussian-based query, larger-better.]{\includegraphics[width=.45\columnwidth]{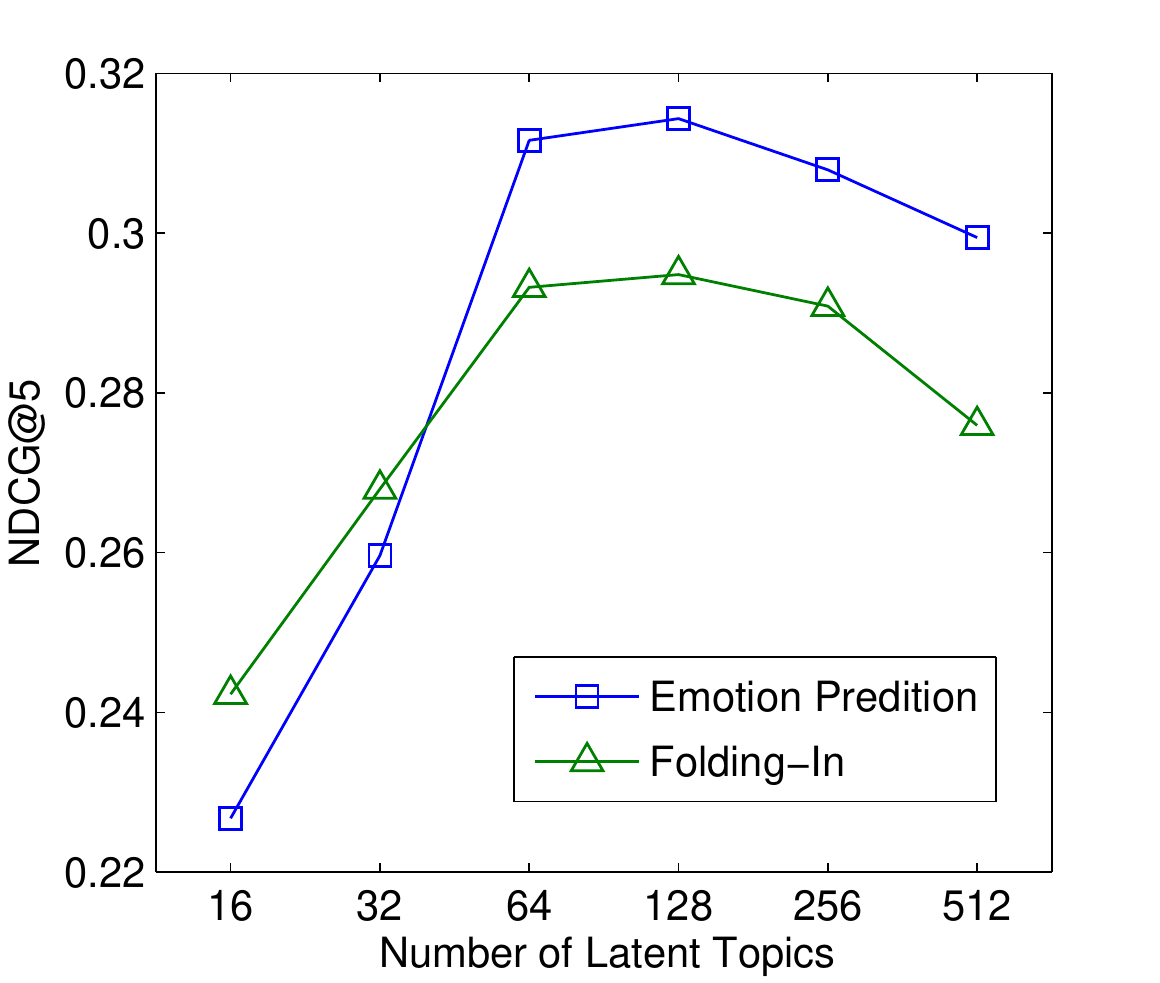}}
\caption{Evaluation result of emotion-based music retrieval.}
\label{fig: EMO_RET}
\end{figure}

Note that we only consider AEG Uniform for simplicity in the result presentation. Our preliminary study reveals that AEG Uniform in general perform slightly better than AEG AnnoPrior and the hybrid method mentioned in Section \ref{sec: GMER_result} in the retrieval task. Moreover, for the Folding-In approach, early stop is not only important to the folding-in process, but also necessary to learning the affective GMM. According to our pilot study, setting $ITER=2-4$ for learning affective GMM and $ITER=3$ for learning the pseudo song lead to the optimal performance.

Figures. \ref{fig: EMO_RET} (a) and (b) compare the NDCG@5 of the Emotion Prediction and Folding-In approaches to emotion-based music retrieval using either point-based or Gaussian-based queries.
We are interested in how the result changes as we vary the number of latent topics.
It can be found that the two approaches perform very similarly for point-based query when $K$ is in between 64 and 256. Moreover, we see that Emotion Prediction can outperform Folding-In for Gaussian-based query when $K$ is sufficiently large ($K\geq 64$). The optimal model is attained when $K=128$ in all cases. Similar to the result in General MER, it seems that setting $K$ either too large or too small would lead to sub-optimal result.

\begin{table}[!t]
\centering
\small
\caption{The query-by-point retrieval performance in terms of NDCG@5, 10, 20, and 30.}{
\begin{tabular}{|l| c c c c|}
\hline
Method  &   $P=5$ & $P=10$ & $P=20$ & $P=30$ \\
\hline\hline
Random	&		0.1398	&	0.1504	&	0.1666	&	0.1804	\\
Emotion Prediction 		&	0.3907	&	0.4027	&	0.4288	&	0.4490	\\
Folding-In 	&		0.3868	&	0.4067	&	0.4333	&	0.4533	\\
Ensemble	&		0.3954	&	0.4129	&	0.4398	&	0.4601	\\
\hline
\end{tabular}}
\label{tab:retreival_res2}
\end{table}

\begin{table}[!t]
\centering
\small
\caption{The query-by-Gaussian retrieval performance in terms of NDCG@5, 10, 20, and 30.}{
\begin{tabular}{|l| c c c c|}
\hline
Method &  $P=5$ & $P=10$ & $P=20$ & $P=30$ \\
\hline\hline
Random	&		0.1032	&	0.1090	&	0.1185	&	0.1272	\\
Emotion Prediction		&	0.3143	&	0.3306	&	0.3481	&	0.3658	\\
Folding-In	&		0.2932	&	0.3147	&	0.3383	&	0.3532	\\
Ensemble	&		0.3204	&	0.3368	&	0.3601	&	0.3783	\\
\hline
\end{tabular}}
\label{tab:retreival_res3}
\end{table}

Tables \ref{tab:retreival_res2} and \ref{tab:retreival_res3} present the result of NDCG@5, 10, 20, and 30 for different retrieval methods, including the random baseline, Emotion Prediction, Folding-In, and the Ensemble approaches. The latter three use AEG Uniform with $K=128$.
It is obvious that the latter three can significantly outperform the random baseline, demonstrating the effectiveness of AEG in emotion-based music retrieval.
It can also be found that the Ensemble approach leads to the best result.

A closer comparison between Emotion Prediction and Folding-In for point-based query shows nip and tuck, whereas the former performs consistently better regardless of the value of $P$ for Gaussian-based query.
Moreover, the NDCG measure seems more favorable for point-based query than Gaussian-based one. Our observation indicates that the standard deviation of the ground truth relevance scores (i.e. $\{R(i)\}_{i=1}^Q$) for Gaussian-based query is much larger, resulting in a more challenging measurement basis than that for point-based query. However, the relative performance difference between the two methods is similar for point-based and Gaussian-based queries.

\section{Conclusion}
\label{sec: CONCLUSION}

AEG is a principled probabilistic framework that nicely unifies the computation processes for MER and emotion-based music retrieval for dimensional emotion representations such as valence and arousal. Moreover, AEG better takes into account the subjective nature of music emotional responses through the use of probabilistic inference and model adaptation, further making it possible to personalize an emotion-based MIR system.
The codes for implementing AEG can be retrieved from the link below: \url{http://slam.iis.sinica.edu.tw/demo/AEG/}.

Despite that AEG is a powerful approach, there remains a number of challenges for MER, including:
\begin{itemize}
\item Is it the best way to consider the valence-arousal space as a coordinate space (with two orthogonal axes)?
\item How do we define the ``intensity" of emotion? Does the magnitude of a point in the emotion space implies intensity?  Would it be possible to train regressors that treat the emotion space as a polar coordinate? 
\item What are the features that are more important for modeling emotion?
\item Cross genre generazability \cite{eerola14aes}.
\item Cross culture generazability \cite{hu14smc}.
\item How to incorporate lyrics features for MER?
\item How to model the effect of the singing voice in emotion perception?
\item How do findings in MER help emotion-based music synthesis or manipulation?
\end{itemize}

We note that AEG is only suitable for an emotion-based MIR system when we characterize emotions in terms of valence and arousal. It does not apply to systems that use categorical mood tags to describe emotion. A corresponding probabilistic model for categorical MER is yet to be developed. More research efforts are also needed for the personalization and retrieval aspects for categorical MER.

The AEG model itself can also be improved in a number of directions.
For example, there are several alternative methods that one can adopt to enhance the latent acoustic descriptors (i.e. $\{A_k\}_{k=1}^K$ in Section \ref{sec: LEARING}) for clip-level topic poster representation, such as deep learning \cite{schmidt13ismir} or sparse representations \cite{Li2014TMM}. One can also perform discriminative training to reduce the prediction error by using the same corpus with respect to the selection of Gaussian components or parameter refinement over the affective GMM. For example, a stacked discriminative learning on the parameters initialized by a EM-learned generative model has been studied for years in speech recognition \cite{juang97tsap,chou03book}. Following this research line, it may help improve AEG as well. Finally, the AEG framework can be easily extended to include multi-modal content such as lyrics, review comments, album cover, and music video. For instance, given a silent video sequence, one can accompany it with a piece of music based on music emotion \cite{wang12grand}.

\section{Acknowledgement}

This work was supported by the Academia Sinica--UCSD Postdoctoral Fellowship to Ju-Chiang Wang, the Academia Sinica Career Development Program to Yi-Hsuan Yang, and the Ministry of Science and Technology of Taiwan under Grants NSC 101-2221-E-
001-019-MY3, NSC 102-2221-E-001-004-MY3, and NSC 102-2221-E-001-008-MY3.

\bibliographystyle{abbrv}
\bibliography{NewBib}

\end{document}